\documentclass{jfm}

\usepackage{graphicx}
\usepackage{natbib}
\usepackage{color}
\usepackage{amsfonts,amssymb,amsmath}
\usepackage{bm}

\graphicspath{{./figures/}}

\ifCUPmtlplainloaded \else
  \checkfont{eurm10}
  \iffontfound
    \IfFileExists{upmath.sty}
      {\typeout{^^JFound AMS Euler Roman fonts on the system,
                   using the 'upmath' package.^^J}%
       \usepackage{upmath}}
      {\typeout{^^JFound AMS Euler Roman fonts on the system, but you
                   dont seem to have the}%
       \typeout{'upmath' package installed. JFM.cls can take advantage
                 of these fonts,^^Jif you use 'upmath' package.^^J}%
      }
  \else
  \fi
\fi

\ifCUPmtlplainloaded \else
  \checkfont{msam10}
  \iffontfound
    \IfFileExists{amssymb.sty}
      {\typeout{^^JFound AMS Symbol fonts on the system, using the
                'amssymb' package.^^J}%
       \usepackage{amssymb}%
         \let\leq=\leqslant
         
      }{}
  \fi
\fi

\ifCUPmtlplainloaded \else
  \IfFileExists{amsbsy.sty}
    {\typeout{^^JFound the 'amsbsy' package on the system, using it.^^J}%
     \usepackage{amsbsy}}
    {\providecommand\boldsymbol[1]{\mbox{\boldmath $##1$}}}
\fi

\newsavebox{\astrutbox}
\sbox{\astrutbox}{\rule[-5pt]{0pt}{20pt}}

\newcommand{\BS}[1]{{\color{red}#1}}

\begin{document}


\title{Linear stability of slip pipe flow}

\author[Kaiwen Chen, Baofang Song]
{Kaiwen Chen, Baofang Song
\thanks{Email address for correspondence: baofang\_song@tju.edu.cn}}
\affiliation{Center for Applied Mathematics, Tianjin University, Tianjin 300072, China 
}

\maketitle

\begin{abstract}
We investigated the linear stability of pipe flow with anisotropic slip length at the wall
by considering streamwise and azimuthal slip separately as the limiting cases.
Our numerical analysis shows that streamwise slip renders the flow less stable but does not cause instability. The exponential decay rate of the least stable mode appears to be $\propto Re^{-1}$ when the Reynolds number is sufficiently large. Azimuthal slip can cause linear instability if the slip length is sufficiently large. The critical Reynolds number can be reduced to a few hundred given large slip lengths. Besides numerical calculations, we present a mathematical proof of the linear stability of the flow to three-dimensional yet streamwise-independent disturbances for arbitrary Reynolds number and slip length, as an alternative to the usual energy analysis. Meanwhile we derived analytical solutions to the eigenvalue and eigenvector, and explained the structure of the spectrum and the dependence of the leading eigenvalue on the slip length. The scaling of the exponential decay rate of streamwise independent modes is shown to be rigorously $\propto Re^{-1}$. Our non-modal analysis shows that overall streamwise slip reduces the non-modal growth and azimuthal slip has the opposite effect. Nevertheless, both slip cases still give the $Re^2$-scaling of the maximum non-modal growth and the most amplified disturbances are still streamwise rolls, which are qualitatively the same as in the no-slip case. 
\end{abstract}


\section{Introduction}{\label{sec:intro}} 
The classic pipe flow with no-slip boundary condition has been proved linearly stable to axisymmetric perturbations \citep{Herron1991,Herron2017}, and numerical studies suggest that the flow is linearly stable to any perturbations at arbitrary Reynolds numbers \citep{Meseguer2003}. The recent work of \citet{Chen2019} presented a rigorous proof of the linear stability of the flow to general perturbations at high Reynolds number regime. Therefore, transition to turbulence in pipe flow is subcritical via finite-amplitude perturbations (see e.g. \citet{Eckhardt2007,Avila2011}).

However, velocity slip of viscous fluid can occur on super-hydrophobic surfaces \citep{Voronov2008,Rothstein2010}, for which slip boundary condition instead of the classic no-slip condition should be adopted for the momentum equations, and the slip boundary condition can potentially influence the stability of the flow. 
A simplified and widely used slip boundary condition is the Navier slip boundary condition, which has been shown to apply to many flow problems and frequently adopted for linear stability studies  \citep[to list a few]{Vinogradova1999, Lauga2005, Min2005, Gan2006, Ren2008, Ghosh2014b, Seo2016, Chattopadhyay2017}. For pipe geometry, although many studies have investigated the linear stability of immiscible and miscible multi-fluid flows with either no-slip or Navier slip boundary condition \citep[etc.]{Hu1989, Joseph1997, Li1999, Selvam2007, Sahu2016, Chattopadhyay2017}, 
much fewer studies were dedicated to the linear stability of single-phase pipe flow with slip boundary condition. \citet{Prusa2009} investigated this problem and showed that, subject to Navier slip boundary condition, pipe flow becomes less stable compared to the no-slip case, however, the destabilization effect is constrained to small Reynolds numbers and is not sufficient to render the flow linearly unstable. Their results indicated that the stability property of pipe flow is not qualitatively affected by the slip boundary condition, regardless of the slip length. For its counterpart in plane geometry, i.e. channel flow, on the contrary, \citet{Min2005,Lauga2005} reported a stabilizing effect of velocity slip on the linear stability.
 
Usually, slip length is assumed homogeneous and isotropic, i.e. independent of position and direction at the wall in stability analysis. However, anisotropy in the effective slip length can be incurred by anisotropy in the texture pattern on superhydrophobic surfaces, such as parallel periodic slats, grooves and grates \citep{Lecoq2004,Bazant2008,Ng2009,Belyaev2010,Asmolov2012,Pralits2017}. For example, \citet{Ng2009} reported a ratio of down to about 0.25 between the transverse slip length (in the direction perpendicular to the slats) and longitudinal slip length (parallel to the slats). The linear stability of channel flow with anisotropic slip caused by parallel micro-graves was analyzed by \citet{Pralits2017} using the the tensorial formulation of slip boundary condition proposed by \citet{Bazant2008}. Their results showed possibilities of linear instability using special alignment of the micro-graves. Recently, \citet{Chai2019} studied the linear stability of single-phase channel flow subject to anisotropy in slip length by considering streamwise and azimuthal slip separately as the limiting cases, which can potentially be realized or approximated by using specially designed surface texture, e.g. specially aligned micro-grates/graves, according to \citet{Bazant2008}. Their results showed that streamwise slip mainly stabilizes the flow (with increased critical Reynolds number), although it surprisingly destabilizes the flow slightly in a small Reynolds number range, and that azimuthal slip can greatly destabilize the flow and reduce the critical Reynolds number given sufficiently large slip length. The critical Reynolds number can be reduced to a few hundred with a dimensionless azimuthal slip length of $\mathcal{O}(0.1)$, in contrast to $Re_{cr}=5772$ for the no-slip case. Their study also indicated that Squire's theorem \citep{Squire1933} ceases to apply when the wall normal velocity and vorticity are coupled via the slip boundary condition, such that the leading instability becomes three dimensional (3-D) rather than two dimensional (2-D) when slip length is sufficiently large, in agreement with \citet{Pralits2017}. The stability of 3-D perturbations was not considered by \citet{Min2005,Lauga2005} in which Squire's theorem was seemingly assumed.  

Differing from channel flow, linear instability is absent at arbitrary Reynolds numbers in classic pipe flow. 
This raises the question of whether the anisotropy in slip length can also cause linear instability in pipe flow. To our knowledge, this problem has not been studied in pipe geometry. 
The pseudospectrum analysis of classic pipe flow of \citet{Schmid1994,Meseguer2003} suggests that, despite the linear stability, at sufficiently large Reynolds numbers, a small perturbation to the linear operator associated with the governing equation can possibly change the stability of the system. The slip boundary condition can be thought of as a perturbation to the linear operator with no-slip boundary condition. However, \citet{Prusa2009} showed that homogeneous and isotropic slip does not change the spectrum qualitatively no matter how large the slip length (i.e. operator perturbation) is. Following \citet{Chai2019}, in this work, we still consider anisotropic slip length in the limiting cases and explore the possibility of linear instability for pipe flow. Aside from the critical Reynolds number as focused on by \citet{Chai2019}, here we also investigate the effects of the slip on the spectrum and on the scaling of the leading eigenvalues with Reynolds number. Besides numerical calculations, we also perform analytical studies on the eigenvalues and eigenvectors of the 3-D yet streamwise-independent modes, and discuss about their structure as well as their dependence on the slip length on a theoretical basis, which to our knowledge have not been reported in the literature.

\section{Numerical methods}\label{sec:methods}

The nondimensional incompressible Navier-Stokes equations read
\begin{equation}\label{equ:NS}
 \dfrac{\partial \bm u}{\partial t}+{\bm u}\cdot\bm{\nabla}
{\bm u}=-{\bm{\nabla}p}+\dfrac{1}{Re}{\bm\nabla^2}{\bm u}, \;
\bm{\nabla}\cdot{\bm u}=0,
\end{equation}
where $\bm u$ denotes velocity and $p$ denotes pressure. For pipe geometry, cylindrical coordinates $(r, \theta, x)$ are considered, where $r$, $\theta$ and $x$ denote the radial, azimuthal, and streamwise coordinates, respectively. Velocity components $u_r$, $u_\theta$ and $u_x$ are normalized by $2U_b$ where $U_b$ is the bulk speed (the average of the streamwise velocity on the pipe cross-section), length by pipe radius $R$ and time by $R/U_b$. The Reynolds number is defined as $Re=U_bR/\nu$ where $\nu$ is the kinematic viscosity of the fluid. In order to eliminate the pressure and impose the incompressibility condition, we adopt the velocity-vorticity formulation of \citet{Schmid1994}, with which the governing equations of disturbances reduce to only two equations about the wall normal velocity $u_r$ and wall normal vorticity $\eta$. With a Fourier-Fourier-Chebyshev collocation discretization, considering perturbations of the form of $\{u_r,\eta\}=\{\hat{u}_r(r), \hat{\eta}(r)\}\text{e}^{-i(\alpha x+n\theta)}$, the governing equations in the Fourier spectral space read
\begin{equation}\label{equ:NS2}
L{\boldsymbol q}+\dfrac{\partial}{\partial \tau}M{\boldsymbol q}=0,
\end{equation}
where 
\begin{equation}\label{equ:operatorL}
L=
\left( \begin{array}{cc}
i\alpha ReU\Gamma +i\dfrac{\alpha Re}{r}\left(\dfrac{U'}{k^2r}\right)'+\Gamma(k^2r^2\Gamma) & 2\alpha n^2Re\Gamma \\
-\dfrac{iU'}{r}+\dfrac{2\alpha}{Re}\Gamma & i\alpha Rek^2r^2U+\phi 
\end{array} 
\right ),
\end{equation}

\begin{equation}\label{equ:operatorM}
M=
\left( \begin{array}{cc}
\Gamma & 0 \\
0 & k^2r^2 
\end{array} 
\right ),
\end{equation}
$\tau=\dfrac{t}{Re}$ is the scaled time, and unknowns are
\begin{equation}\label{def:q}
\boldsymbol q=
\left(\begin{array}{c}
\hat \Phi \\ 
\hat\Omega
\end{array}
\right)=
\left(\begin{array}{c}
-ir{\hat u_r} \\ 
\dfrac{\alpha r{\hat u_\theta}-n{\hat u_x}}{nRek^2r^2}
\end{array}
\right)=
\left(\begin{array}{c}
-ir{\hat u_r} \\ 
\dfrac{\hat \eta}{inRek^2r}
\end{array}
\right).
\end{equation}
The real number $\alpha$ is the axial wave number and $n$, which is an integer, is the azimuthal wavenumber. The base flow is denoted as $U$, $k^2=\alpha^2+\dfrac{n^2}{r^2}$, $i=\sqrt{-1}$ and the prime denotes the derivative with respect to $r$. The operators $\Gamma$ and $\phi$ are defined as $\Gamma=\dfrac{1}{r^2}-\dfrac{1}{r}\dfrac{\text{d}}{\text{d}r}\left(\dfrac{1}{k^2r}\dfrac{\text{d}}{\text{d}r}\right)$ and 
$\phi=k^4r^2-\dfrac{1}{r}\dfrac{\text{d}}{\text{d}r}\left(k^2r^3\dfrac{\text{d}}{\text{d}r} \right)$.
The other two velocity components $\hat u_{x}$ and $\hat u_{\theta}$ can be calculated as 
\begin{equation}
\hat u_x=-\dfrac{\alpha}{k^2r}\dfrac{\partial \hat{\Phi}}{\partial r}-n^2r\hat\Omega,\hspace{4mm} \hat u_{\theta}=-\dfrac{n}{k^2r^2}\dfrac{\partial \hat{\Phi}}{\partial r}+\alpha nrRe\hat{\Omega}.
\end{equation}

We use the Robin-type Navier slip boundary condition at the pipe wall for streamwise and azimuthal velocities separately, i.e.
\begin{equation}\label{equ:BC}
\left(l_x\dfrac{\partial{u_x}}{\partial r}+u_x\right)\bigg|_{r=1}=0, \hspace{2mm} \left(l_\theta\dfrac{\partial{u_\theta}}{\partial r}+u_\theta\right)\bigg|_{r=1}=0, 
\end{equation}
where $l_x\geqslant 0$ and $l_\theta\geqslant 0$ are streamwise and azimuthal slip lengths, respectively, and are independent of each other. In spectral space, these boundary conditions apply identically to $\hat u_{x}$ and $\hat u_\theta$ given the homogeneity of the slip length. We use the no-penetration condition for the wall-normal velocity component at the wall, i.e. $u_r(1,\theta,x,t)=0$. \citet{Lauga2005, Chai2019} considered the same boundary conditions for slip channel flow. Note that in the isotropic slip case considered by \citet{Prusa2009}, $l_x$ and $l_\theta$ are related as $l_\theta=\dfrac{l_x}{1+l_x}$, which gives $l_\theta\approx l_x$ for small slip lengths. With boundary condition (\ref{equ:BC}), given that we impose the same volume flux as in the no-slip case, i.e.
\begin{equation}\label{equ:mass_flux}
\int_0^1U_x(r)r{\text d}r=\frac{1}{4},
\end{equation}
the velocity profile of the constant-volume-flux base flow reads
\begin{equation}\label{equ:base_flow}
\boldsymbol U(r)=\dfrac{1-r^2+2l_x}{1+4l_x}\hat{\boldsymbol x},
\end{equation}
where $\hat{\boldsymbol x}$ represents the unit vector in the streamwise direction. Note that the base flow is independent of $l_\theta$. 
Converting to the $(\hat\Omega,\hat\Phi)$ system, the boundary condition (\ref{equ:BC}) reads
\begin{equation}\label{equ:BC2_1}
\dfrac{\alpha}{k^2}\dfrac{\partial \hat{\Phi}}{\partial r}+n^2Re{\hat \Omega}+l_x\left(n^2Re\dfrac{\partial \hat \Omega}{\partial r}+\dfrac{\alpha}{k^2}\dfrac{\partial^2{\hat \Phi}}{\partial r^2}+\alpha\dfrac{n^2-\alpha^2}{(n^2+\alpha^2)^2}\dfrac{\partial {\hat \Phi}}{\partial r}\right)=0
\end{equation} 
and
\begin{align}\label{equ:BC2_2}
\alpha nRe{\hat \Omega} &-\dfrac{n}{n^2+\alpha^2}\dfrac{\partial {\hat \Phi}}{\partial r}+
\notag \\
& l_\theta\left( \alpha nRe{\hat \Omega}+\alpha nRe\dfrac{\partial{\hat \Omega}}{\partial r} -\dfrac{n}{n^2+\alpha^2}\dfrac{\partial^2{\hat \Phi}}{\partial r^2}+\dfrac{2n\alpha^2}{(n^2 +\alpha^2)^2}\dfrac{\partial {\hat \Phi}}{\partial r}\right)=0.
\end{align} 
It can be seen that $\hat\Omega$ and $\hat\Phi$, i.e. $\hat u_r$ and $\hat\eta$, are coupled via the slip boundary condition.

In order to avoid the singularity at the pipe center, i.e. $r=0$, the domain [0, 1] is extended to [-1, 1] and an even number of Chebyshev grid points over [-1, 1] are used such that there is no grid point at $r=0$. This extension also allows us to use the Chebyshev collocation method for the discretization in the radial direction and the resulted redundancy is circumvented by setting proper parity conditions on $\hat\Phi$ and $\hat\Omega$ with respect to $r$ \citep{Trefethen2000,Meseguer2003}. In this way, no explicit boundary condition is imposed at the pipe center.

To determine whether a mode $(\alpha,n)$ is linearly stable or not, one only needs to calculate the eigenvalues of the operator $-M^{-1}L$ and check if any eigenvalue has a positive real part, $\lambda_r$, which determines the asymptotic growth/decay rate of the corresponding eigenvector as $t\to\infty$. 

\section{Streamwise slip}\label{sec:streamwise_slip}
We consider the case of $l_x\neq 0$ and $l_\theta=0$ 
as the limiting case of streamwise slip being significant and azimuthal slip being negligible.

\begin{figure}
\centering
\includegraphics[width=0.95\textwidth]{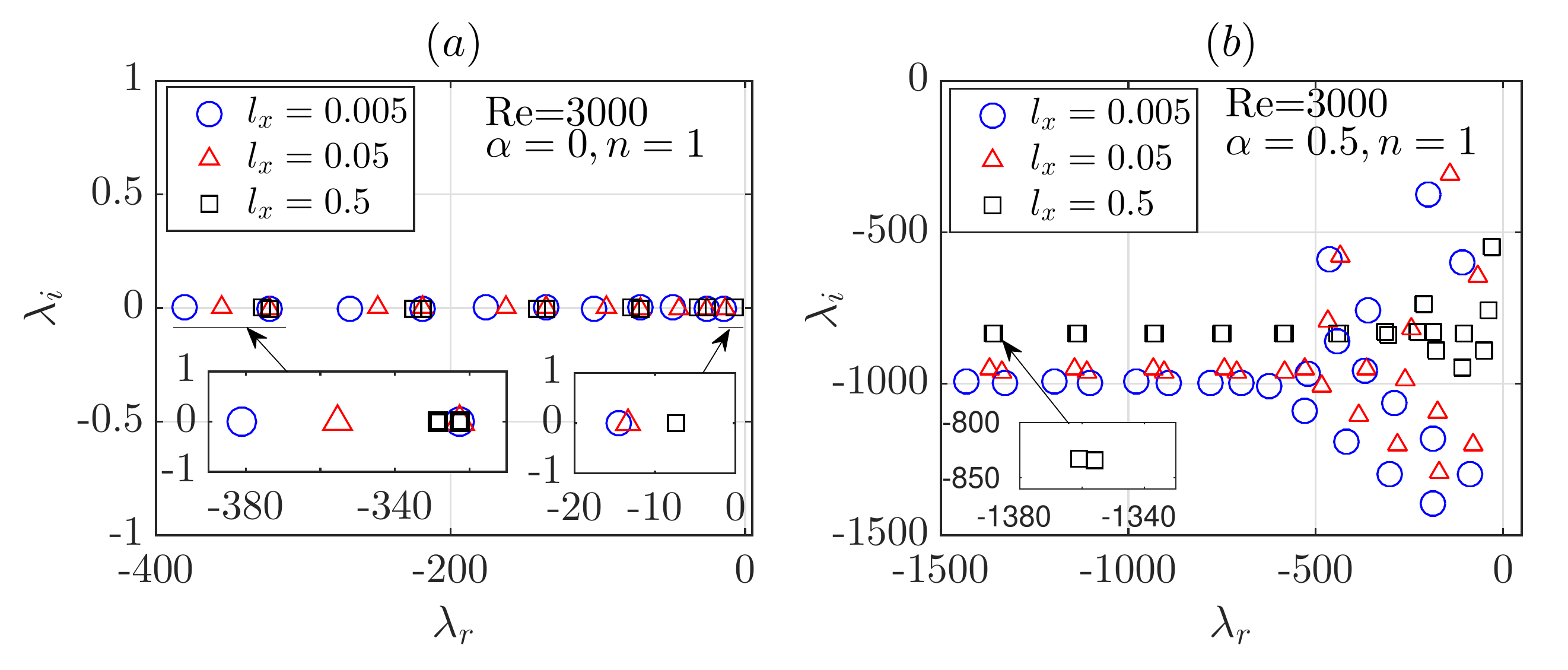}
\caption{\label{fig:spectrum_streamwise_slip} Spectrum of the flow at $Re=3000$ with $l_x=0.005$ (circles), 0.05 (triangles) and 0.5 (squares). (a) The mode $(\alpha,n)=(0,1)$. (b) The mode $(\alpha,n)=(0.5,1)$.}
\end{figure}

The effect of the slip on the spectrum is investigated for $Re=3000$ and is shown in Figure \ref{fig:spectrum_streamwise_slip} for the modes $(\alpha,n)=(0,1)$ and $(0.5,1)$. Firstly, panel (a) shows that the eigenvalues of the $(\alpha,n)=(0,1)$ mode visually all fall on the $\lambda_i=0$ line ($\lambda_i$ denotes the imaginary part of the eigenvalue) and in the left half-plane, which suggests that the eigenvalues are all real and negative. \citet{Meseguer2003} reported the same finding for the no-slip case in a large Reynolds number range up to $10^7$. In fact, the eigenvalues being real and negative can be rigorously proved, see our proof in Section \ref{sec:proof_linear_stability_alpha0}.
Secondly, as $l_x$ increases, it can be observed that there are two groups of eigenvalue, one of which stays constant and the other of which shifts to the right, see the two insets in panel (a). Specifically, as $l_x$ is increased to 0.5, the left eigenvalue in the left inset has moved from the circle to the triangle and finally to the square while the right eigenvalue stays constant. Nontheless, the rightmost eigenvalue increases as $l_x$ increases (see the right inset) which indicates that the flow becomes less stable. In Section \ref{sec:eigenvalue_eigenvector_alpha0}, we will show that the former group corresponds to disturbances with $\Phi\not\equiv 0$, i.e. $u_r\not\equiv 0$ and the latter group, on the contrary, is associated with disturbances with $\Phi\equiv 0$, i.e. $u_r\equiv 0$ and, the rightmost eigenvalue belongs to the latter group (see Figure \ref{fig:eigenvalue_validation}).
Panel (b) shows the case for the mode $(\alpha,n)=(0.5,1)$. The slip does not qualitatively change the shape of the spectrum. As $l_x$ increases, the eigenvalues overall move to the right. Besides a horizontal shift, there is a shift in the vertical direction either, and meanwhile the spectrum is compressed in the vertical direction, see the comparison between the $l_x=0.5$ and the other two cases. Using the term of \citet{Schmid1994,Meseguer2003}, the horizontal branch of the spectrum (the part with $\lambda_r\lesssim -600$) corresponds to mean modes, the upper branch corresponds to wall modes and the lower branch to center modes. Note that the speed of a wave is given by $\frac{-\lambda_i}{\alpha Re}$ in our formulation. It has been known that the wave speed of the mean modes follows the mean velocity of the `two-dimensional' axial base flow, i.e. $\int_0^1U_x(r)\text{d}r$ in pipe flow (see e.g. \citet{Drazin1981}), which gives $\frac{2}{3}$ in the no-slip case \citep{Schmid1994}. In our case, the wave speed of the mean modes is decreased by the slip, reducing to 0.5559 for $l_x=0.5$ ($\frac{833.868}{0.5\times 3000}$, see the eigenvalue in Table \ref{tab:streamwise_slip}) which is very close to $\frac{5}{9}$ given by $\int_0^1U_x(r)\text{d}r$ with the base flow shown in (\ref{equ:base_flow}). The wall modes, which are located close to the wall, move at lower speed than the center modes, which are located close to the pipe center and move at speeds close to the centerline velocity.  Since we fix the volume flux of the flow while the slip length is varied, the speed of the base flow close to the wall increases as $l_x$ increases, whereas the speed near the pipe center decreases, i.e. the velocity profile becomes flatter, see the base flow given by (\ref{equ:base_flow}). Therefore, it can be expected that as $l_x$ increases, the speed of the wall-modes increases and that of the center modes decreases, and all three types of modes move at closer speeds. This is exactly what the compression in the vertical direction of the spectrum reveals. 
The other noticeable effect is that the slip brings the adjacent eigenvalues associated with the mean modes closer as the slip length increases, causing a seemingly degeneracy of the spectrum, see Figure \ref{fig:spectrum_streamwise_slip}(b).

\begin{figure}
\centering
\includegraphics[width=0.9\textwidth]{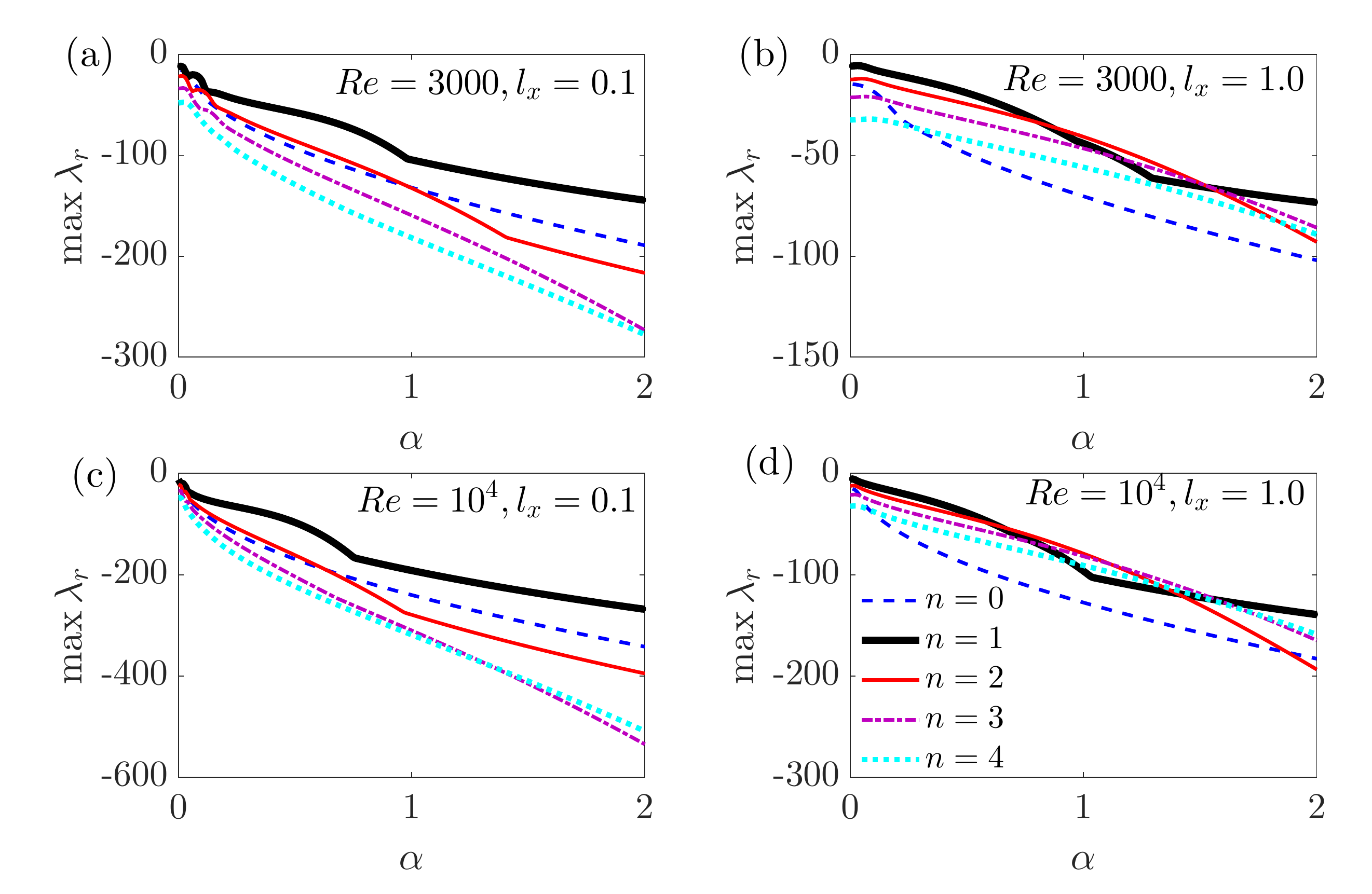}
\caption{\label{fig:Re3000_10000_lx0.1_1.0} The maximum eigenvalue, $\max{\lambda_r}$, as a function of $\alpha$, for $Re=3000$ (a,b) and $10^4$ (c,d). For each Reynolds number, azimuthal wavenumbers $n=0$, 1, 2, 3, 4 and slip lengths $l_x=0.1$ and 1.0 are shown.}
\end{figure}

Figure \ref{fig:Re3000_10000_lx0.1_1.0} shows the maximum of the real part of the eigenvalue, $\max{\lambda_r}$, as a function of the streamwise wavenumber, $\alpha$, for $Re=3000$ and $10^4$. For each $Re$, slip lengths $l_x=0.1$ and 1.0, and azimuthal wavenumbers $n=0$, 1, 2, 3 and 4 are considered. The trend shown in the figure suggests that, for both Reynolds numbers, $\alpha=0$ is nearly the least stable mode, i.e. the slowest decaying mode given that all $\max{\lambda_r}$'s are negative, regardless of the slip length. At small $\alpha$, where $\max{\lambda_r}$ is largest, the results suggest that $n=1$ is always the least stable one. At larger $\alpha$, however, $n=1$ is still the least stable when $l_x$ is small, see the case of $l_x=0.1$ in panel (a, c), but is not in a range of $\alpha$ around $\alpha=1$, see the case of $l_x=1.0$ in panel (b, d). Nevertheless, in this range, $\max{\lambda_r}$ is much smaller than that in the small $\alpha$ regime. Therefore,  as we are most interested in the least stable mode, in the following, we will focus on the $n=1$ modes.
In fact, for the $\alpha=0$ modes, we can rigorously prove that $n=1$ is the least stable azimuthal wavenumber, see Appendix \ref{sec:dependence_on_n_alpha0}. 

\begin{figure}
\centering
\includegraphics[width=0.85\textwidth]{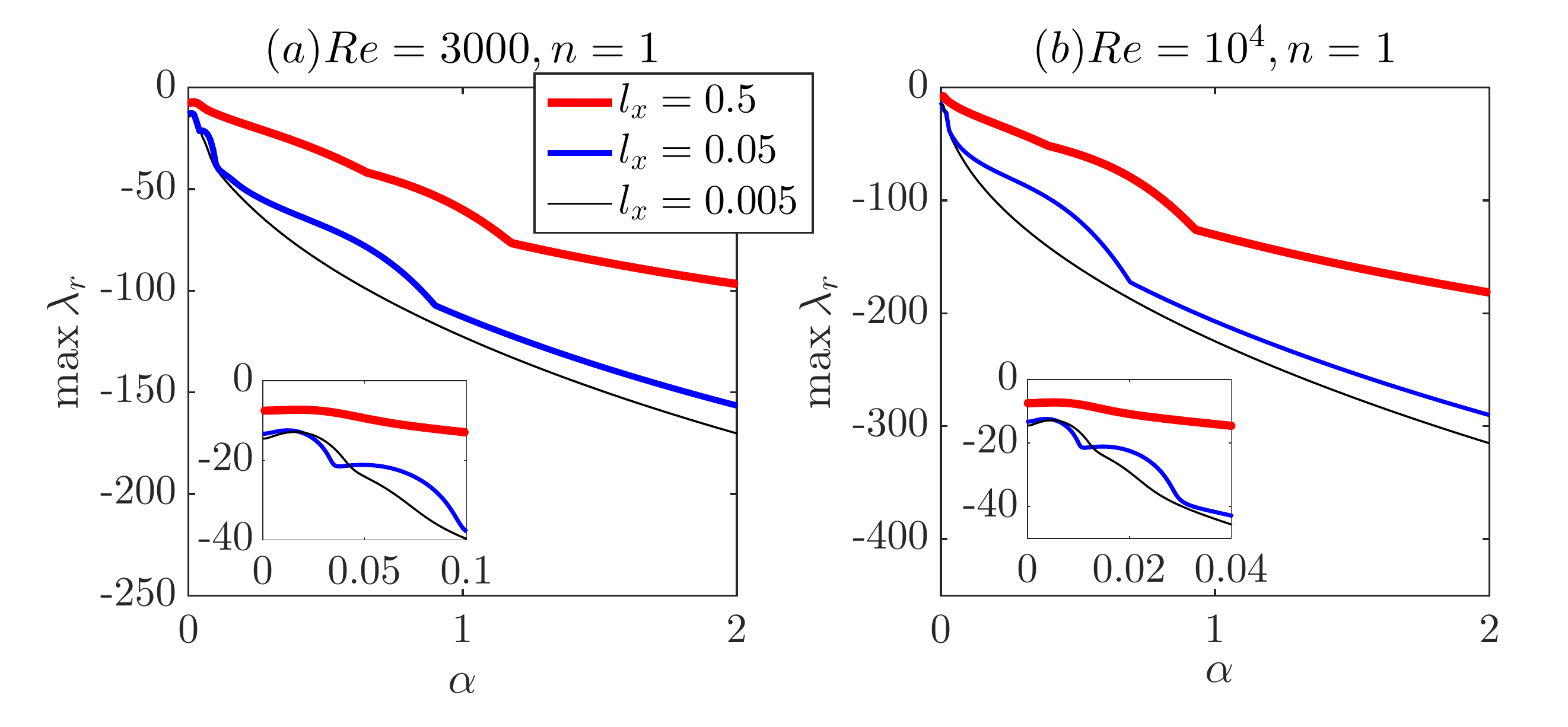}
\caption{\label{fig:Re3000_10000_n1_alpha_dependence} The influence of streamwise slip on $\max{\lambda_r}$ of $n=1$ modes for $Re=3000$ (a) and $10^4$ (b). Slip lengths of $l_x=0.005$ (thin black), 0.05 (blue) and 0.5 (bold red) are shown. The insets show the close-up of the regions with very small $\alpha$.}
\end{figure}

Figure \ref{fig:Re3000_10000_n1_alpha_dependence} shows $\max{\lambda_r}$ as a function of $\alpha$ of the $n=1$ modes for $Re=3000$ (a) and $10^4$ (b). For each $Re$, overall $\max{\lambda_r}$ increases as $l_x$ increases, i.e. the $n=1$ modes decay more slowly as $l_x$ increases. The insets show the close-up of the small $\alpha$ region, in which the dependence of $\max{\lambda_r}$ on $\alpha$ is not monotonic, with the maximum appears at some small but finite $\alpha$ instead of $\alpha=0$. Nevertheless, the difference between the peak value and the value for $\alpha=0$ is very small, i.e. $\alpha=0$ is nearly the least stable mode, as aforementioned. In fact, the dependence on $l_x$ is not fully monotonic either, see the very small region around $\alpha=0.03$ for the $l_x=0.005$ (the thin black line) and $l_x=0.05$ (\BS{the blue line}) cases as shown in the inset in (a) and around $\alpha=0.01$ in the inset in (b). However, for $\alpha=0$ and in most range of $\alpha$, our results show a monotonic increase of $\max{\lambda_r}$ as $l_x$ increases. 

\begin{figure}
\centering
\includegraphics[width=0.85\textwidth]{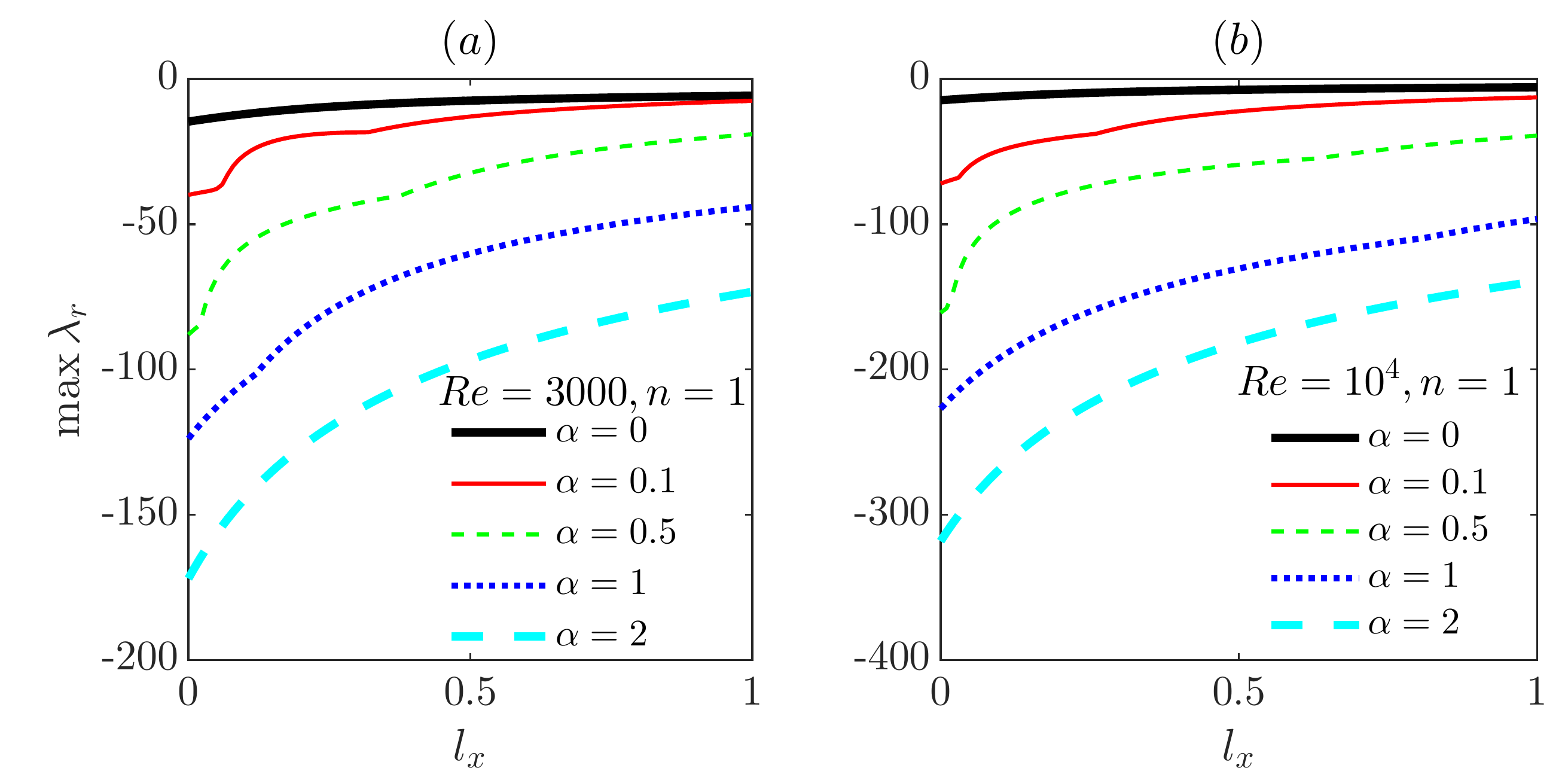}
\caption{\label{fig:Re3000_10000_n1_a_few_alpha} $\max{\lambda_r}$ of $n=1$ modes with $\alpha=0$, 0.1, 0.5, 1.0 and 2.0 as a function of $l_x$ for $Re=3000$ (a) and $Re=10^4$ (b).}
\end{figure}

Figure \ref{fig:Re3000_10000_n1_a_few_alpha} illustrates the dependence of $\max{\lambda_r}$ of the $n=1$ modes on $l_x$ in a broader range of $l_x$. For each $Re$, $\alpha=0$, 0.1, 0.5, 1 and 2 are shown. The trend shows that as $l_x$ keeps increasing, $\max{\lambda_r}$ seems to asymptotically approach a plateau with a negative value, i.e. all the modes shown in the figure appear to be linearly stable, for both Reynolds numbers. 

\begin{figure}
\centering
\includegraphics[width=0.85\textwidth]{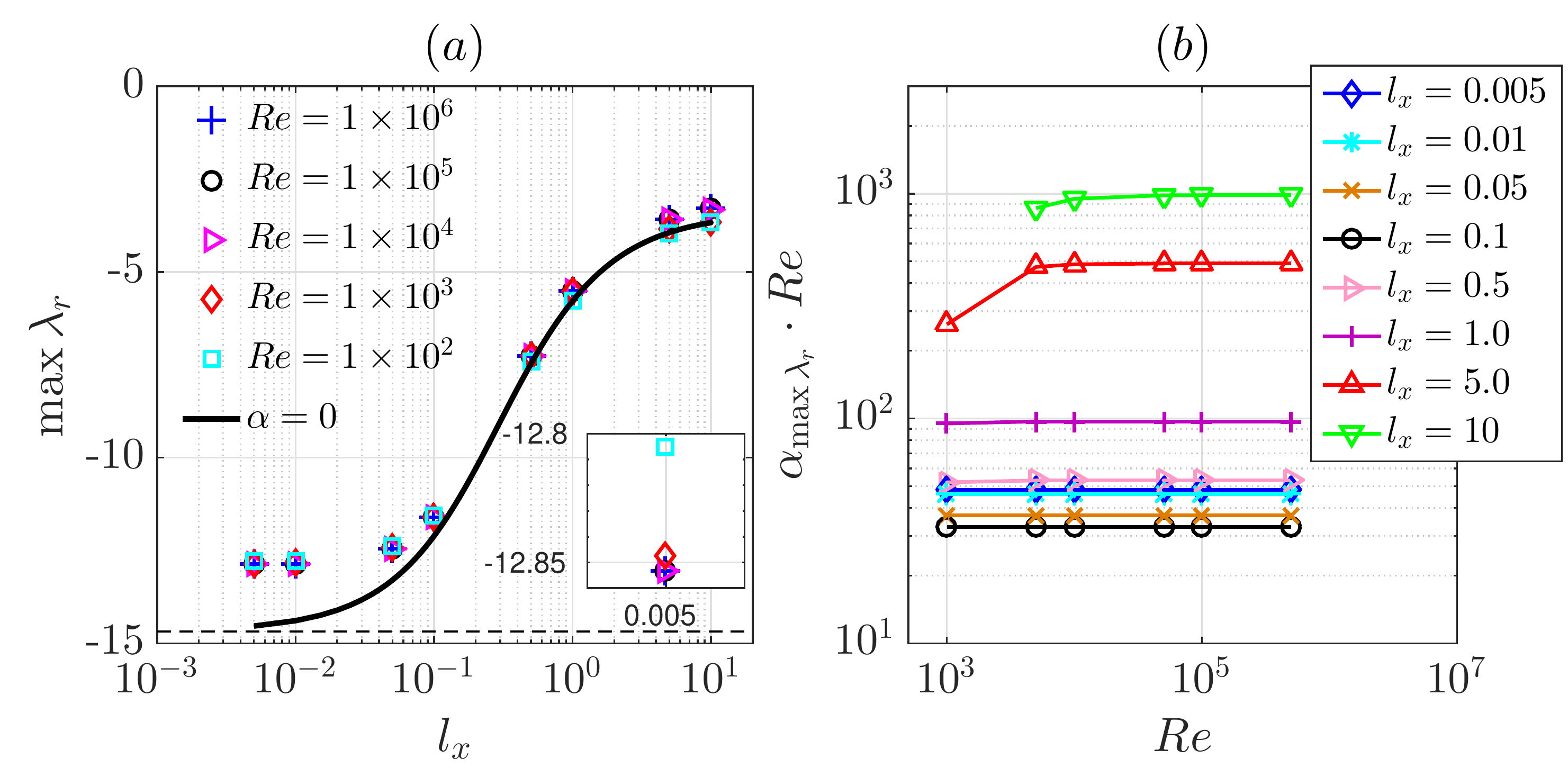}
\caption{\label{fig:max_eigenvalue_vs_slip_up_to_Re1000000} (a) The global maximum of $\max{\lambda_r}$, i.e. the maximum of $\max{\lambda_r}$ over $\alpha$ and $n$, for $Re=100$, 1000, $1\times 10^4$, $1\times 10^5$, and $1\times 10^6$ (symbols). The bold black line shows the maximum $\max{\lambda_r}$ of the $\alpha=0$ modes, which is associated with the $(\alpha, n)=(0, 1)$ mode and is independent of $Re$. The dashed line marks the value for the $(\alpha, n)=(0, 1)$ mode in the no-slip case \citep{Meseguer2003}. The inset shows the zoom-in at $l_x=0.005$. (b) The product of $Re$ and $\alpha_{\max{\lambda_r}}$ (the $\alpha$ at which $\max{\lambda_r}$ takes the maximum) plotted against $Re$.} 
\end{figure}

The above results suggest that, with streamwise slip, the flow is linearly stable to any perturbations, regardless of the slip length. In order to show evidences in a broader parameter regime, we numerically searched for the global maximum of $\max{\lambda_r}$ over $\alpha$ and $n$ and explored a wider range of $l_x$ up to 10 and of $Re$ up to $10^6$. Practically, based on our analysis, we only need to search in a small range of $\alpha$ immediately above zero (see the insets in Figure \ref{fig:Re3000_10000_n1_alpha_dependence}) while setting $n=1$. Specifically, the range of [0, 1.2] is searched at $Re=100$, and the range is decreased as $Re^{-1}$ for higher Reynolds numbers.
Then we plotted the global maximum of $\max{\lambda_r}$, still denoted as $\max{\lambda_r}$, as a function of $l_x$, for a few Reynolds numbers ranging from $10^2$ to $10^6$ in Figure \ref{fig:max_eigenvalue_vs_slip_up_to_Re1000000}(a). 

It is interesting to note that our data for high Reynolds numbers all collapse over the whole $l_x$ range investigated, see the cases with $Re$ above $1\times 10^4$ in Figure \ref{fig:max_eigenvalue_vs_slip_up_to_Re1000000}, suggesting that the maximum eigenvalue of the system is independent of $Re$. At lower Reynolds numbers, e.g. $Re=100$ and $10^3$ in the figure, there is almost a collapse for small $l_x$ ($\lesssim 0.1$) but a small deviation from the high Reynolds number cases can be seen, see the inset that shows the zoom-in at $l_x=0.005$. As $l_x$ increases further,
the maximum eigenvalue for $Re=100$ and $10^3$ approaches to that of $\alpha=0$ modes, which is strictly $Re$-independent (see the proof in Section \ref{sec:proof_linear_stability_alpha0}). Besides, the figure also shows that the global maximum of $\max{\lambda_r}$ is slightly larger than the maximum of the $\alpha=0$ modes over the whole $l_x$ range and the difference is most significant at small $l_x$. We did not explore further larger $l_x$ considering that the range we investigated is already much larger than the slip length that can be encountered in applications ($\lesssim 0.1$ in set-ups with characteristic length of one millimeter or larger, because so far the maximum slip length achieved in experiments is $\mathcal{O}(100)$ micron, see \citet{Voronov2008,Lee2008, Lee2009}). Nevertheless, the $S$-shaped trend as $l_x$ increases suggests that the flow stays stable no matter how large the slip length is. In fact, as $l_x\to\infty$, full slip boundary condition is recovered, and the velocity profile of the base flow will be completely flat and no mean shear exists, in which case linear stability can be expected for any perturbations. Panel (b) shows the product of $Re$ and the $\alpha$ at which $\max{\lambda_r}$ maximizes globally, denoted as $\alpha_{\max{\lambda_r}}$. Interestingly, it seems that this product is a constant when $l_x$ is small ($\lesssim 0.1$) for all the $Re$'s investigated and approaches a constant as $Re$ is sufficiently high ($\gtrsim 10^4$) if $l_x\gtrsim 0.1$. This indicates that $\alpha_{\max{\lambda_r}}$ scales as $Re^{-1}$ for either not very large $l_x$ or in high Reynolds number regime. It should be noted that we observed a non-monotonic dependence of $\alpha_{\max{\lambda_r}}\cdot Re$ on the slip length, which minimizes at around $l_x=0.1$.

That the global $\max{\lambda_r}$ is $Re$-independent, as our results suggest, indicates that the slowest exponential decay rate (referred to as decay rate for simplicity hereafter) of perturbations scales as $Re^{-1}$ given that the scaled time $\tau=\frac{t}{Re}$ is used in our formulation, see Eqs. (\ref{equ:NS2}). The same scaling was observed by the calculation of \citet{Meseguer2003} for the $(\alpha,n)=(0,1)$ mode of the classic pipe flow. Therefore, our results suggest that, as $Re\to \infty$, the decay rate of perturbations asymptotically approaches zero and stays negative, i.e. the flow is linearly stable at arbitrary Reynolds number. The $Re^{-1}$-scaling of the slowest decay rate can be rigorously proved for the $\alpha=0$ modes, see Section \ref{sec:proof_linear_stability_alpha0}. 

In a word, in the pure streamwise slip case, we did not observe any linear instability in the large ranges of $l_x$ and $Re$ that we considered, and based on the data shown in Figure \ref{fig:max_eigenvalue_vs_slip_up_to_Re1000000}, we propose that streamwise slip destabilizes the flow but does not cause linear instability, regardless of the slip length and Reynolds number. A similar destabilizing effect was reported by \citet{Prusa2009} for the isotropic slip case.

\section{Azimuthal slip}\label{sec:azimuthal_slip}
We consider the case of $l_\theta\neq 0$ and $l_x=0$ as the limiting case of azimuthal slip being significant and streamwise slip being negligible.

\begin{figure}
\centering
\includegraphics[width=0.95\textwidth]{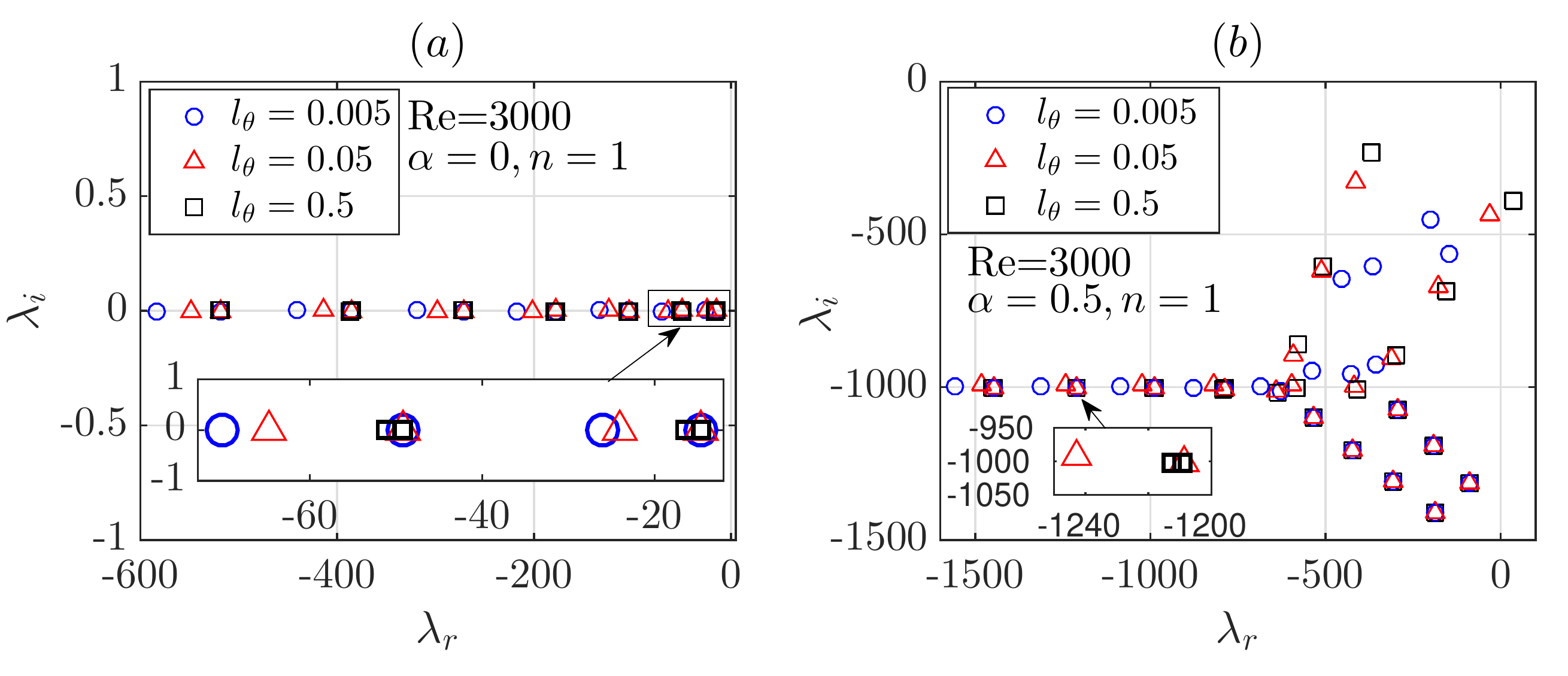}
\caption{\label{fig:spectrum_spanwise_slip} Spectrum of the flow at $Re=3000$ with $l_\theta=0.005$ (circles), 0.05 (triangles) and 0.5 (squares). (a) The mode $(\alpha,n)=(0,1)$. (b) The mode $(\alpha,n)=(0.5,1)$.}
\end{figure}

The effect of azimuthal slip on the spectrum is investigated for $Re=3000$ and is shown in Figure \ref{fig:spectrum_spanwise_slip} for the modes $(\alpha,n)=(0,1)$ and $(0.5,1)$. Similar to the streamwise slip case, the eigenvalues of the $\alpha=0$ mode also fall on the $\lambda_i=0$ line and in the left half-plane, see panel (a). This suggests that the eigenvalues of streamwise-independent modes are all real and negative. We will show a rigorous proof of this observation in Section \ref{sec:proof_linear_stability_alpha0}. As $l_\theta$ increases, 
similar to the streamwise slip case, we also observed two groups of eigenvalues. One group stays constant as the azimuthal slip length changes and the other shifts to the right, see the inset in panel (a). 
As we will theoretically show in Section \ref{sec:eigenvalue_eigenvector_alpha0} and \ref{sec:dependence_on_slip_length_alpha0}, the former group is associated with the disturbances with $\Phi\equiv 0$ and is independent of $l_\theta$, and the latter group is associated with the disturbances with $\Phi\not\equiv 0$. The rightmost eigenvalue belongs to the former group for $l_\theta<1$ and can only be overtaken by the latter group if $l_\theta>1$ (the two groups precisely overlap when $l_\theta=1$), i.e. the rightmost eigenvalue can only increase with $l_\theta$ if $l_\theta>1$.
For the $\alpha=0.5$ and $n=1$ mode, the mean-mode branch overall does not show either a vertical or horizontal shift, but adjacent eigenvalues are brought closer by the increasing slip length, and for $l_\theta=0.5$ there is almost an eigenvalue degeneracy (see the inset in panel (b)). The center-mode branch is nearly unchanged as $l_\theta$ increases. However, the wall-mode branch is significantly affected. As $l_\theta$ increases, the wall mode overtakes the center mode and becomes the least stable perturbation, and as $l_\theta$ is sufficiently large, the rightmost eigenvalue appears in the right half-plane, indicating the onset of a linear instability. 
In contrast to the streamwise slip case, the wave speed of the mean modes does not change because the speed follows $\int_0^1U_x(r)\text{d}r$ as aforementioned and the base flow $\boldsymbol U(r)$ is not affected by the azimuthal slip. The speed of the center modes is also not affected, whereas the wave speed of the wall modes is considerably decreased by the slip. This is reasonable because the slip boundary condition should mostly affect the flow close to the wall and should not affect significantly the flow far from the wall. 

\begin{figure}
\centering
\includegraphics[width=0.4\textwidth]{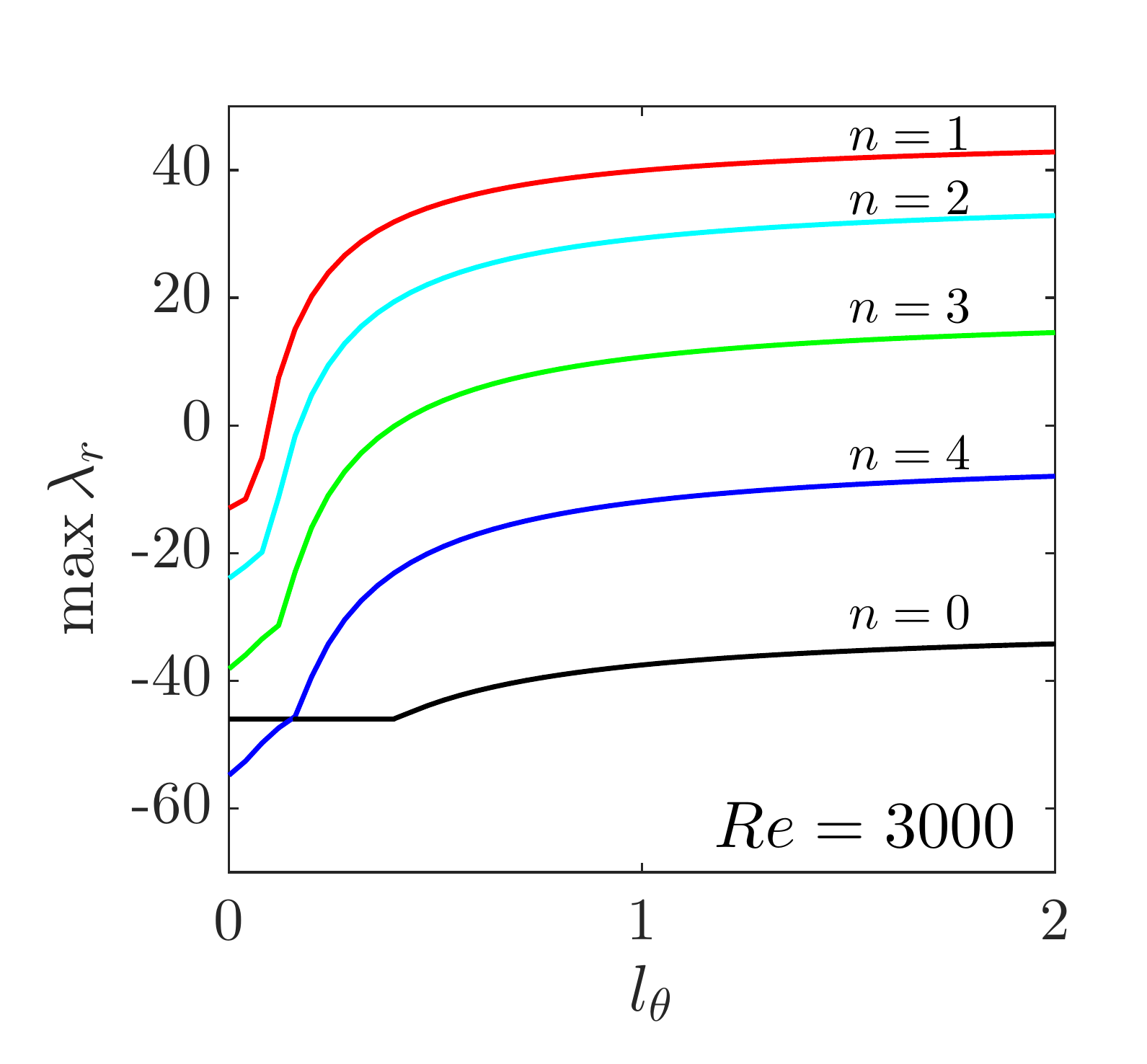}
\caption{\label{fig:maxeig_vs_lambdat_n0to4_Re3000} The $\max{\lambda_r}$ maximized over $\alpha$, still denoted as $\max{\lambda_r}$, as a function of $l_\theta$. Modes with $n=0$, 1, 2, 3 and 4 are shown for $Re=3000$.}
\end{figure}

Figure \ref{fig:maxeig_vs_lambdat_n0to4_Re3000} shows $\max{\lambda_r}$ maximized over $\alpha$ (over [0, 2] in practice), still denoted as $\max{\lambda_r}$, as a function of $l_\theta$ for $n=0$, 1, 2, 3 and 4 at $Re=3000$. Overall, $\max{\lambda_r}$ increases monotonically as $l_\theta$ increases, while the $n=0$ case seems to stay constant until it starts to increase at around $l_\theta=0.4$. In the small $l_\theta$ regime, all modes are linearly stable. As $l_\theta$ is increased to around 0.1, $\max{\lambda_r}$ of the $n=1$ mode becomes positive, indicating a linear instability. As $l_\theta$ increases further, $n=2$ and 3 also become unstable. In the whole range of $l_\theta$ investigated, $n=1$ is the least stable/most unstable one, which is also the case for other Reynolds numbers we investigated. Therefore, in the following, we mainly discuss about $n=1$ modes.

\begin{figure}
\centering
\includegraphics[width=0.8\textwidth]{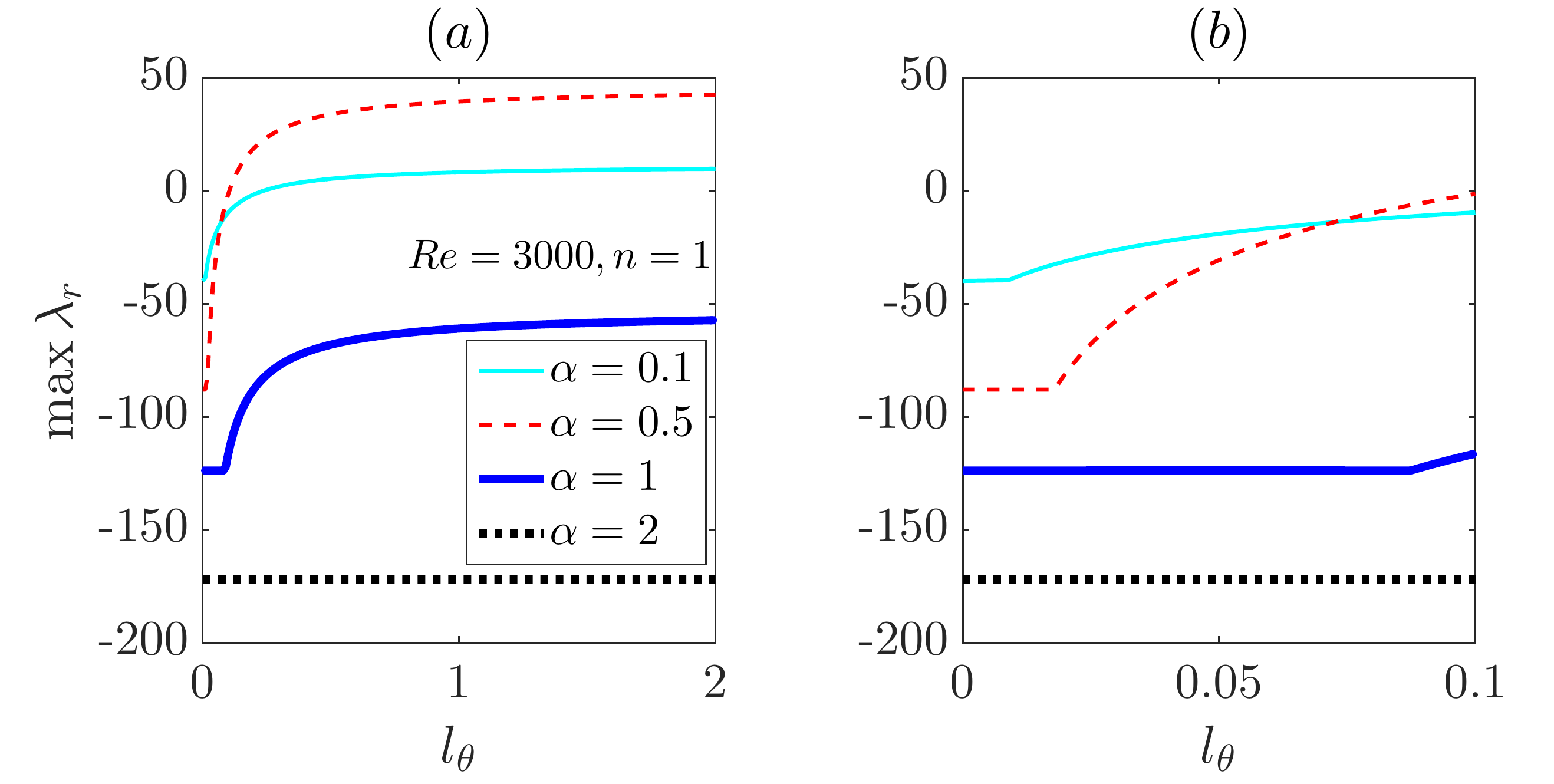}
\caption{\label{fig:maxeig_vs_lambdat_a_few_alpha_Re3000} $\max{\lambda_r}$ of modes $\alpha=0.1$, 0.5, 1.0 and 2.0 for $Re=3000$ and $n=1$ as a function of $l_\theta$. Panel (b) shows the details in the small $l_\theta$ regime.}
\end{figure}

Figure \ref{fig:maxeig_vs_lambdat_a_few_alpha_Re3000} shows $\max{\lambda_r}$ of modes $\alpha=0.1$, 0.5, 1.0 and 2.0 for $Re=3000$ and $n=1$ as a function of $l_\theta$. The results show that when $l_\theta$ is small, overall $\max{\lambda_r}$ decreases as $\alpha$ increases. As $l_\theta$ is increased, some moderate $\alpha$ turns to be the least stable/most unstable mode, see the crossover of $\alpha=0.1$ (cyan thin line) and 0.5 (red dashed line) cases in the figure. Panel (b) shows the small $l_\theta$ range, in which it appears that $\max{\lambda_r}$ first stays nearly unchanged and then starts to increase, and the trend shows that the larger $\alpha$, the later $\max{\lambda_r}$ starts to increase as $l_\theta$ is increased. The same behavior is also observed for $\alpha=0$ modes and we will show a rigorous proof of this behavior in Section \ref{sec:dependence_on_slip_length_alpha0}. Interestingly, the case of $\alpha=2$ seems to stay unchanged up to $l_\theta=2.0$.

\begin{figure}
\centering
\includegraphics[width=0.45\textwidth]{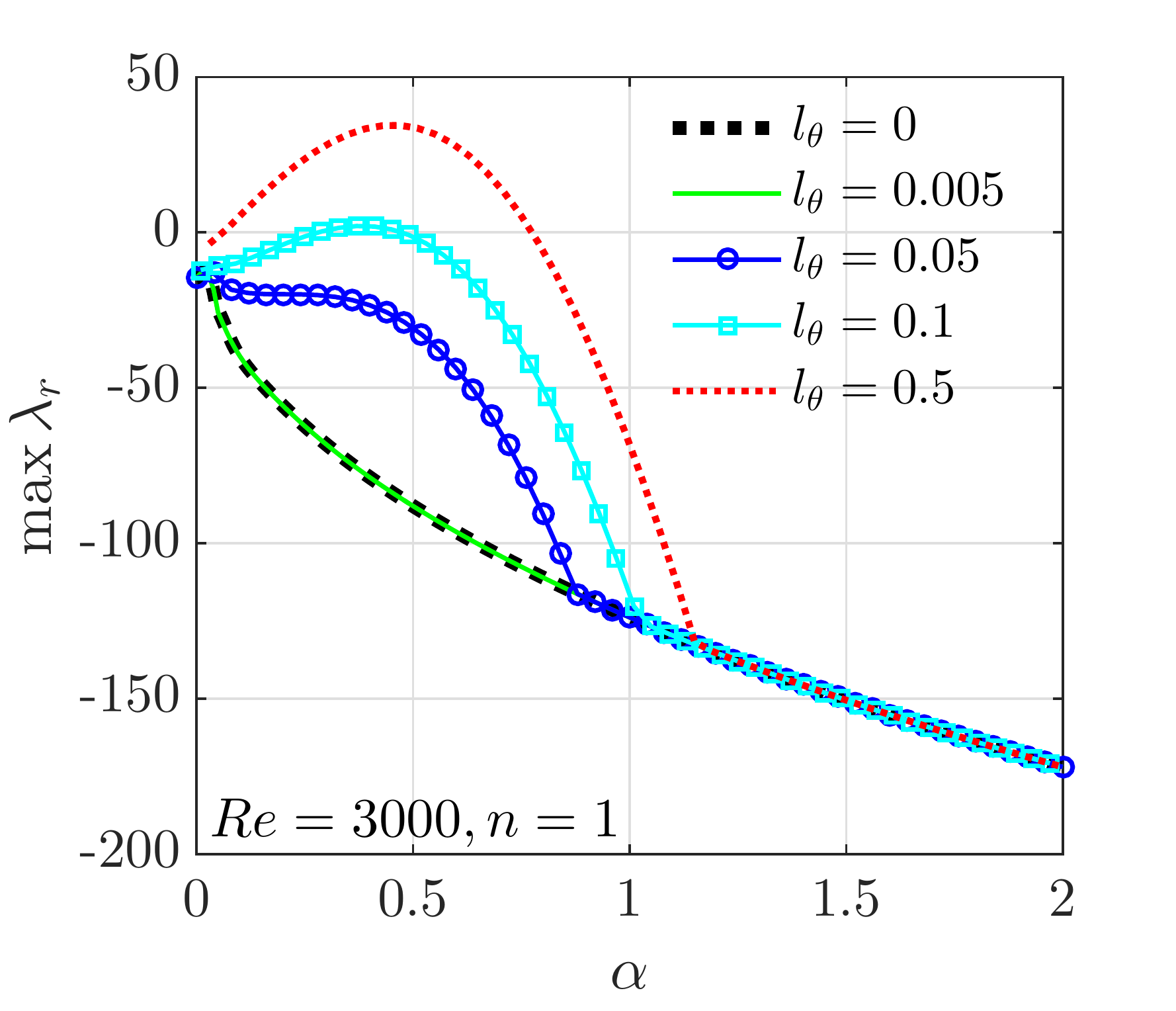}
\caption{\label{fig:maxeig_vs_alpha_a_few_lambdat_Re3000} $\max{\lambda_r}$ of the $n=1$ modes as a function of $\alpha$ for $Re=3000$. The data for $l_\theta=0$, 0.005, 0.05, 0.1 and 0.5 are plotted. Note that the curves for $l_\theta=0$ (the black bold dotted line) and $l_\theta=0.005$ (the green thin solid line) coincide.}
\end{figure}

The dependence of $\max{\lambda_r}$ on $\alpha$ is more comprehensively shown in Figure \ref{fig:maxeig_vs_alpha_a_few_lambdat_Re3000}. The smallest $l_\theta=0.005$ shows a monotonic decrease with increasing $\alpha$, which completely collapses onto the curve for $l_\theta=0$, i.e. the classic no-slip case. However, as $l_\theta$ increases, $\max{\lambda_r}$ significantly increases in the region of $0<\alpha\lesssim 1$ such that a bump appears in the curves, see those for $l_\theta=0.05$, 0.1 and 0.5. At certain point, the bump reaches $\max{\lambda_r=0}$ and the flow starts to become linearly unstable if $l_\theta$ increases further, see the cases of $l_\theta=0.1$ and 0.5. As observed in Figure \ref{fig:maxeig_vs_lambdat_a_few_alpha_Re3000} for the $\alpha=2$ case, the results suggest that $\max{\lambda_r}$ of sufficiently large $\alpha$ seems unaffected by azimuthal slip in the $l_\theta$ range investigated, see the collapse of all curves above $\alpha\simeq 1.2$ in Figure \ref{fig:maxeig_vs_alpha_a_few_lambdat_Re3000}. It should be noted that as $l_\theta$ becomes larger, the bump widens up, i.e. $\max{\lambda_r}$ is affected by the slip in a wider range of $\alpha$. 

\begin{figure}
\centering
\includegraphics[width=0.8\textwidth]{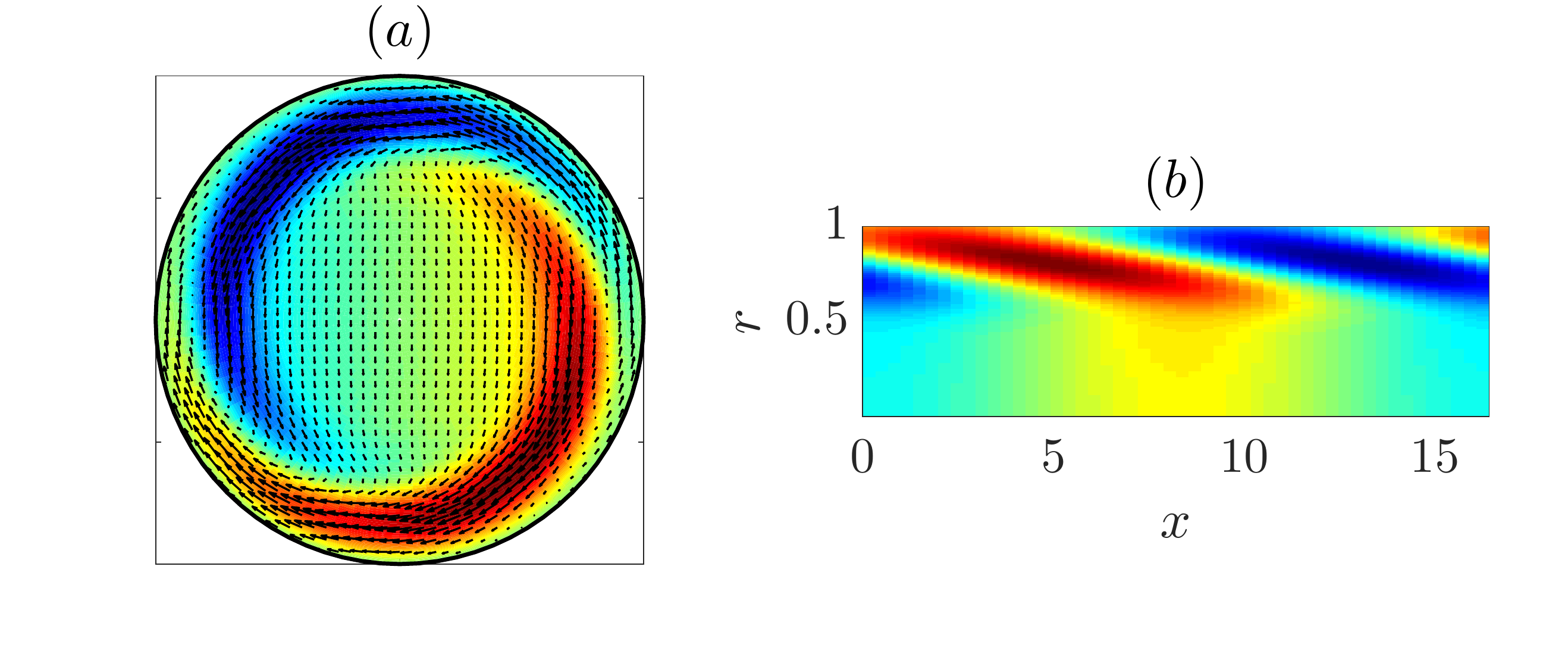}
\caption{\label{fig:most_unstable_mode_Re3000} Visualization of the most unstable mode $(\alpha,n)=(0.383,1)$ for $Re=3000$ with $l_\theta=0.1$. Panel (a) shows the $r-\theta$ cross-section and (b) the $x-r$ cross-section. In both panels, the streamwise velocity is plotted as the colormap with red color representing positive and blue representing negative values with respect to the base flow. In (a), the in-plane velocity field is plotted as arrows.}
\end{figure}

Figure \ref{fig:most_unstable_mode_Re3000} shows the velocity field of the most unstable perturbation of mode $(\alpha,n)=(0.383, 1)$ for $Re=3000$ with $l_\theta=0.1$. Panel (a) shows the in-plane velocity field in the $r-\theta$ pipe cross-section and (b) shows the pattern of the streamwise velocity in the $x-r$ cross-section. The patterns shown suggest that the flow manifests with a pair of helical waves. The flow structures are mostly located near the wall ($r\gtrsim 0.5$), indicating that the most unstable mode is a wall-mode, which can also been seen in Figure \ref{fig:spectrum_spanwise_slip}(b). 

\begin{figure}
\centering
\includegraphics[width=0.9\textwidth]{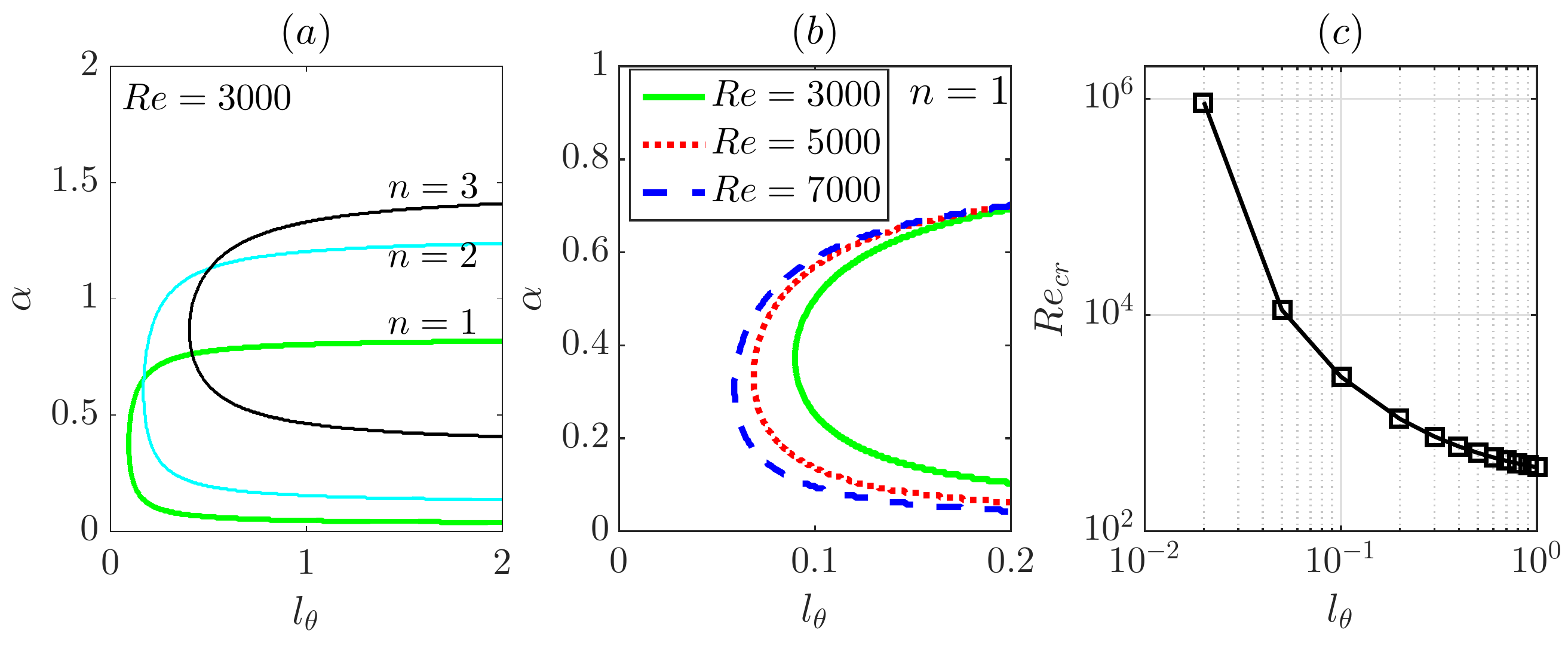}
\caption{\label{fig:stability_boundary_critical_Reynolds_number}(a)The neutral stability curves for $Re=3000$ and $n=1$, 2 and 3 in the $l_\theta$-$\alpha$ plane. (a) The neutral stability curves for $n=1$ and $Re=3000$, 5000 and 7000. (b) The critical Reynolds number $Re_{cr}$ as a function of $l_\theta$.}
\end{figure}

Obviously, azimuthal slip can cause linear instability given sufficiently large slip length.
We can search in the $l_\theta$-$\alpha$ plane to obtain the neutral stability curve for given $Re$ and $n$. 
Figure \ref{fig:stability_boundary_critical_Reynolds_number}(a) shows the neutral stability curves for $Re=3000$ and $n=1$, 2 and 3 ($n=4$ and higher are all stable, see Figure \ref{fig:maxeig_vs_lambdat_n0to4_Re3000}). It can be seen that, as $n$ increases, the unstable region shifts to the right and upward. However, as the results in Figure \ref{fig:maxeig_vs_lambdat_n0to4_Re3000} show, $n=1$ is the most unstable based on the eigenvalue maximized over $\alpha$ and therefore, we only investigate the $n=1$ case in the following.
Panel (b) shows the neutral stability curves for $n=1$ and $Re=3000$, 5000 and 7000. As $Re$ increases, the neutral stability curve moves towards the smaller $l_\theta$ region, indicating that, for a given $l_\theta$, the flow becomes more unstable as $Re$ increases, as expected. 
The data show that the wavelength of the unstable modes is comparable or significantly larger than the pipe diameter, whereas very long waves ($\alpha\to 0$) and short waves ($\alpha\gg 1.0$) are generally stable.
That the flow is always stable to perturbations with $\alpha=0$, regardless of the value of $l_\theta$ and $Re$, can be rigorously proved (see Section \ref{sec:proof_linear_stability_alpha0}). 

\begin{figure}
\centering
\includegraphics[width=0.75\textwidth]{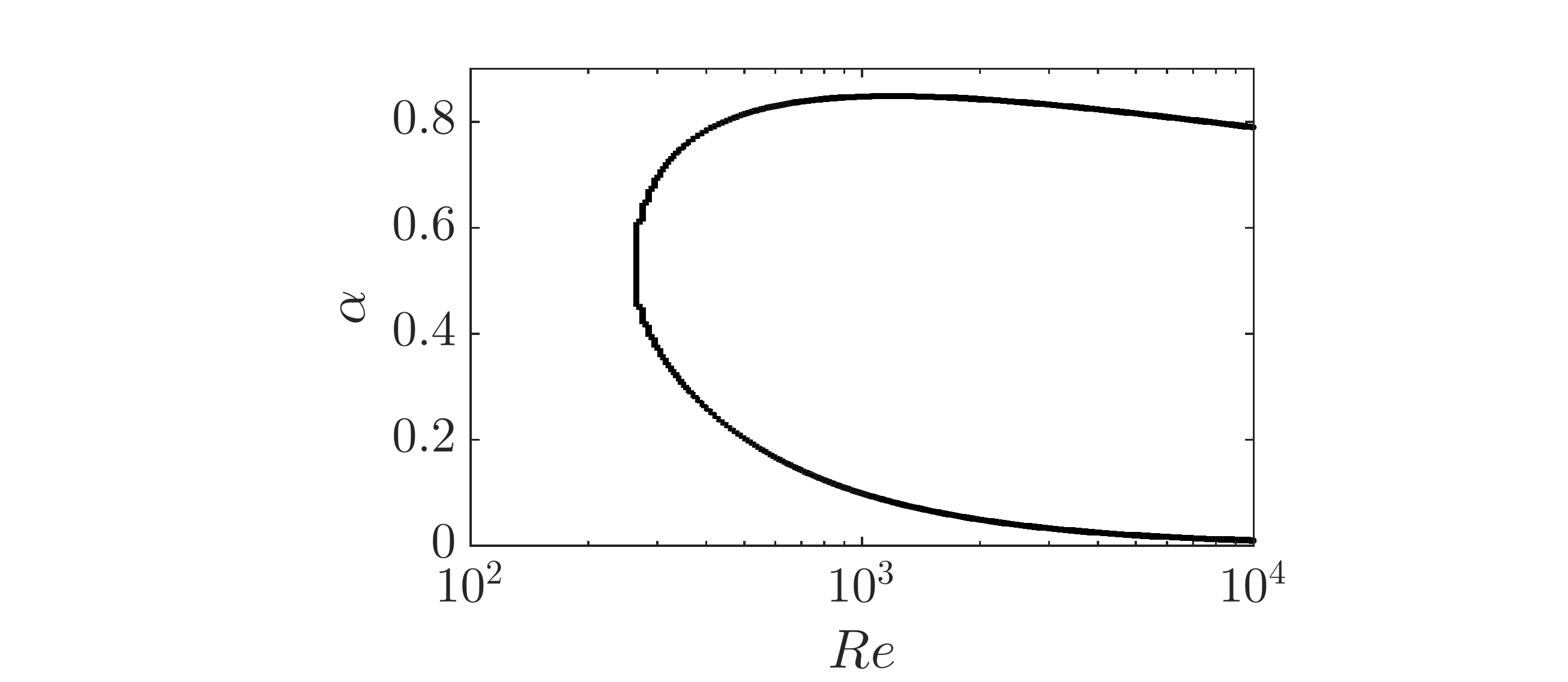}
\caption{\label{fig:stability_boundary_lambdat_infinity} The neutral stability curve in the $Re-\alpha$ plane for $l_\theta=\infty$ and $n=1$.}
\end{figure}

Further, for each $l_\theta$, a critical Reynolds number $Re_{cr}$ can be determined by searching the first appearance of a positive $\max{\lambda_r}$ in the $l_\theta$-$\alpha$ plane by varying $Re$. Figure \ref{fig:stability_boundary_critical_Reynolds_number}(c) shows $Re_{cr}$ as a function of $l_\theta$. As shown, $Re_{cr}$ is a few hundred if $l_\theta$ is large ($l_\theta\gtrsim 0.3$), but the trend suggests that it does not reduce to zero if $l_\theta\to\infty$. Since the classic pipe flow is linearly stable for arbitrary Reynolds number, there is an explosive increase in $Re_{cr}$ as $l_\theta$ decreases, which can be expected because the classic pipe flow will be recovered if $l_\theta\to 0$.
We also explored the limit of $l_\theta\to\infty$, in which case the boundary condition for the azimuthal velocity becomes the full slip condition of 
\begin{equation}
\dfrac{\partial u_\theta}{\partial r}=0. 
\end{equation}
The neutral stability curve for $n=1$ in the $Re-\alpha$ plane is shown in Figure \ref{fig:stability_boundary_lambdat_infinity}, which shows that the unstable modes are still long waves with $\alpha$ approximately between 0 and 0.8. The critical Reynolds number (the nose of the curve) appears approximately at $Re_{\text{cr}}=260$.

\section{Eigenvalues and eigenvectors of streamwise independent modes }

We can rigorously prove the linear stability of the base flow to perturbations with $\alpha=0$. In the following, we do not consider the $(\alpha,n)=(0, 0)$ mode, which should be strictly stable as it is purely dissipative and there can be no energy production mechanism associated with it. In fact, the stability of the classic pipe flow to streamwise independent perturbations has already been proved by \citet{Joseph1971} using an energy analysis. Nevertheless, here we also account for the effect of the velocity slip and perform analytical studies on the eigenvalues and eigenvectors of $\alpha=0$ modes. 

\subsection{Proof of linear stability to $\alpha=0$ modes}
\label{sec:proof_linear_stability_alpha0}
For $\alpha=0$, the eigenvalue $\lambda$ of the operator $-M^{-1}L$ satisfies
\begin{equation}\label{equ:eigen_equation_Phi}
\Gamma(n^2\Gamma)\Phi+\lambda\Gamma\Phi=0,
\end{equation}
and
\begin{equation}\label{equ:eigen_equation_Omega}
\dfrac{2i}{1+4l_x}\Phi+\phi\Omega+\lambda n^2\Omega=0,
\end{equation}
where $\Phi$ and $\Omega$ compose the eigenvector $\boldsymbol q$ associated with $\lambda$ (see the definition of $q$ in (\ref{def:q})).
The boundary conditions (\ref{equ:BC2_1}) and (\ref{equ:BC2_2}) reduce to
\begin{equation}\label{equ:BCx_alpha0}
l_x\Omega'+\Omega=0
\end{equation}
and
\begin{equation}\label{equ:BCt_alpha0}
l_\theta\Phi''+\Phi'=0
\end{equation}
It can be seen that for $\alpha=0$ modes, $\Omega$ and $\Phi$ are decoupled in the boundary conditions (\ref{equ:BCx_alpha0}) and (\ref{equ:BCt_alpha0}).

We define a space $\Theta=\{f|f\in C^2[0, 1], f(0)=f(1)=0\}$ and an inner product associated with this space
\begin{equation}\label{equ:inner_product}
(f_1,f_2)=\int_0^1rf_1\bar f_2dr,
\end{equation}
where the over-bar represents complex conjugate.
Then the operator $\Gamma$ has the following two properties.
\begin{enumerate}
 \item 
\begin{equation}\label{equ:gamma_p1}
(\Gamma f_1, f_2)=(f_1, \Gamma f_2),\hspace{2mm} \forall f_1, f_2\in \Theta,
\end{equation}

\begin{proof}\label{proof_gamma_p1}
\begin{align}
(\Gamma f_1, f_2)=&\int_0^1r\left( \dfrac{f_1}{r^2}-\dfrac{1}{r}\dfrac{\text{d}}{\text{d}r}\left(\dfrac{r}{n^2}\dfrac{\text{d}f_1}{\text{d}r}\right)\right)\bar f_2\text{d}r \notag\\
=&\int_0^1\dfrac{f_1\bar f_2}{r}\text{d}r-\int_0^1\bar f_2\text{d}\left(\dfrac{r}{n^2}\dfrac{\text{d}f_1}{\text{d}r}\right) \notag\\
=&\int_0^1\dfrac{f_1\bar f_2}{r}\text{d}r-\bar f_2\left( \dfrac{r}{n^2}\dfrac{\text{d}f_1}{\text{d}r}\right)\bigg|_0^1+\int_0^1\dfrac{r}{n^2}\dfrac{\text{d}f_1}{\text{d}r}\text{d}\bar f_2 \notag\\
=&\int_0^1\dfrac{f_1\bar f_2}{r}\text{d}r + \int_0^1\dfrac{r}{n^2}\dfrac{\text{d}\bar f_2}{\text{d}r}\text{d} f_1
\end{align}
and similarly, using intergration by parts, it can be derived that
\begin{align}
(f_1, \Gamma f_2)=&\int_0^1r\left( \dfrac{\bar f_2}{r^2}-\dfrac{1}{r}\dfrac{\text{d}}{\text{d}r}\left(\dfrac{r}{n^2}\dfrac{\text{d}\bar f_2}{\text{d}r}\right)\right)f_1\text{d}r \notag\\
=&\int_0^1\dfrac{f_1\bar f_2}{r}\text{d}r + \int_0^1\dfrac{r}{n^2}\dfrac{\text{d}\bar f_2}{\text{d}r}\text{d} f_1=(\Gamma f_1, f_2)
\end{align}
\end{proof} 
\item 
\begin{equation}\label{equ:gamma_p2}
(\Gamma f, f)\geqslant 0,\hspace{2mm} \forall f\in\Theta.
\end{equation}

\begin{proof}
Taking $f=f_1=f_2$ in Proof \ref{proof_gamma_p1},
\begin{align}
(\Gamma f, f)=\int_0^1\dfrac{f\bar f}{r}\text{d}r + \int_0^1\dfrac{r}{n^2}\dfrac{\text{d}\bar f}{\text{d}r}\text{d}f=\int_0^1\dfrac{f\bar f}{r}\text{d}r+\int_0^1\dfrac{r}{n^2}\dfrac{\text{d}\bar f}{\text{d}r}\dfrac{\text{d}f}{\text{d}r}\text{d}r\geqslant 0.
\end{align}
\end{proof}
Note that property (\ref{equ:gamma_p2}) still holds for those $f$ with $f(1)\neq 0$ but satisfy $f(1)+bf'(1)=0$, where $b>0$ is a constant, because
\begin{align}
(\Gamma f, f)&=\int_0^1\dfrac{f\bar f}{r}\text{d}r-f\left(\dfrac{r}{n^2}\dfrac{\text{d}\bar f}{\text{d}r}\right)\bigg|_0^1+\int_0^1\dfrac{r}{n^2}\dfrac{\text{d}\bar f}{\text{d}r}\text{d}f \notag \\
& =\int_0^1\dfrac{f\bar f}{r}\text{d}r+\int_0^1\dfrac{r}{n^2}\dfrac{\text{d}\bar f}{\text{d}r}\dfrac{\text{d}f}{\text{d}r}\text{d}r + \dfrac{1}{b n^2}f(1)\bar{f}(1)\geqslant0.
\end{align}
\end{enumerate}

Firstly, for the case of $\Phi\equiv 0$ (i.e. the wall normal velocity component $u_r\equiv 0$) and $\Omega\not\equiv 0$, 
Eqs. (\ref{equ:eigen_equation_Omega}) becomes 
\begin{equation}\label{equ:Phi_always_0_1}
\phi\Omega+\lambda n^2\Omega=0,
\end{equation}
and the operators $\phi$ and $\Gamma$ are related as $\phi=\dfrac{n^4}{r^2}-\dfrac{1}{r}\dfrac{\text{d}}{\text{d}r}\left(n^2r\dfrac{\text{d}}{\text{d}r}\right)=n^4\Gamma$. Therefore, Eqs. (\ref{equ:Phi_always_0_1}) becomes
\begin{equation}\label{equ:Phi_always_0_2}
n^4\Gamma\Omega+\lambda n^2\Omega=0.
\end{equation}
Taking the inner product of Eqs. (\ref{equ:Phi_always_0_2}) with $\Omega$, we have
\begin{equation}
(n^4\Gamma\Omega, \Omega)+(\lambda n^2\Omega, \Omega)=0.
\end{equation}
According to property (\ref{equ:gamma_p2}), $(n^4\Gamma\Omega, \Omega)\geqslant 0$ given $\Omega(1)=0$ (without streamwise slip) or $\Omega(1)+l_x \Omega'(1)=0$ (with streamwise slip), which leads to $\lambda<0$, i.e. the eigenvalue is real and negative.

Secondly, we discuss about the $\Phi\not\equiv 0$ case, i.e. the wall normal velocity component $u_r\not\equiv 0$. From Eqs. (\ref{equ:eigen_equation_Phi}), by denoting $g=n^2\Gamma\Phi+\lambda \Phi$, we have $\Gamma g=0$,  i.e.
\begin{equation}
n^2g=r(rg')',
\end{equation}
from which it can be obtained that
\begin{equation}
r^ng=Cr^{2n}+C_1,
\end{equation}
where $C$ and $C_1$ are constants. Note that for $n=2$, $r^2g=n^2r^2\Gamma\Phi+\lambda r^2\Phi=n^2\Phi-nr(r\Phi')'$ has to vanish at $r=0$, because $\Phi$ vanishes, and $\Phi'$ and $\Phi''$ are finite at $r=0$. The same applies to $n>2$. If $n=1$, $rg=\dfrac{\Phi}{r}-(r\Phi')'+\lambda r\Phi=\dfrac{\Phi}{r}-\Phi'-r\Phi''+\lambda r\Phi$, which also vanishes when $r\to 0$ (using L'Hopital rule). Therefore, $C_1\equiv 0$ and $r^ng =Cr^{2n}$, i.e.
\begin{align}\label{equ:rhs_Crn}
n^2\Gamma \Phi + \lambda\Phi  =Cr^n.
\end{align}
\subsubsection{The case without azimuthal slip, i.e. $l_\theta=0$}
\label{sec:stability_alpha0_without_aimuthal_slip}
In case of $l_\theta=0$, the boundary condition (\ref{equ:BC2_2}) or (\ref{equ:BCt_alpha0}) becomes $\Phi'=0$. Taking the inner product (\ref{equ:inner_product}) of Eqs. (\ref{equ:rhs_Crn}) and $\Gamma \Phi$, we have
\begin{equation}\label{equ:eigen_value_alpha0}
n^2(\Gamma\Phi, \Gamma\Phi)+\lambda(\Phi,\Gamma\Phi)=C(r^n,\Gamma\Phi)=C(\Gamma r^n, \Phi)=C(0, \Phi)=0.
\end{equation}
The second equality in Eqs. (\ref{equ:eigen_value_alpha0}) holds in spite of that $r^n\notin \Theta$ and thus, property (\ref{equ:gamma_p1}) cannot be applied directly. Nevertheless, as $\Phi=\Phi'=0$ at $r=1$ in case of $l_\theta=0$, property (\ref{equ:gamma_p1}) still holds (this can been seen by taking $\Phi$ as $f_2$ and $r^n$ as $f_1$ in Proof \ref{proof_gamma_p1}). What follows is that the eigenvalue $\lambda$ is real and $\lambda < 0$ because $(\Gamma \Phi, \Gamma\Phi)>0$ and $(\Phi, \Gamma\Phi)>0$.

\subsubsection{The case with azimuthal slip, i.e. $l_\theta\neq 0$}
\label{sec:stability_alpha0_with_aimuthal_slip}
In case of $l_\theta\neq 0$, Eqs. (\ref{equ:eigen_value_alpha0}) does not hold, except for $C=0$, because $\Phi'=0$ at $r=1$ does not necessarily hold and therefore the second equality in Eqs. (\ref{equ:eigen_value_alpha0}) does not hold either. For $C\neq 0$, consider the special case of $C=1$ (if $C\neq 1$, a rescaling of $\tilde \Phi=\Phi/C$ can easily convert to this special case), Eqs. (\ref{equ:rhs_Crn}) can be written as
\begin{equation}\label{equ:azimuthal_slip_C_1}
(n^2+\lambda r^2)\Phi -r(r\Phi')'=(n^2+\lambda r^2)\Phi -r(\Phi'+r\Phi'')=r^{n+2}.
\end{equation}
As $r\to 1$, Eqs. (\ref{equ:azimuthal_slip_C_1}) turns to
\begin{equation}\label{equ:azimuthal_slip_at_wall_equation_Phi}
-(\Phi'+\Phi'')=1.
\end{equation}
Further, the azimuthal slip requires
\begin{equation}\label{equ:azimuthal_slip_at_wall_BC_Phi}
\Phi'(1)+l_\theta \Phi''(1)=0, \hspace{2mm} l_\theta\in(0,+\infty).
\end{equation}
It follows that, for $l_\theta=1$, $C$ has to be zero, otherwise Eqs. (\ref{equ:azimuthal_slip_at_wall_equation_Phi}) and (\ref{equ:azimuthal_slip_at_wall_BC_Phi}) would conflict with each other. That $C=0$ leads to $\lambda<0$, see Eqs. (\ref{equ:eigen_value_alpha0}).
For $l_\theta\neq 1$, one can solve for $\Phi'$ from Eqs. (\ref{equ:azimuthal_slip_at_wall_equation_Phi}) and (\ref{equ:azimuthal_slip_at_wall_BC_Phi}) as
\begin{equation}\label{equ:Phi_prime_at_wall}
\Phi'(1)=\dfrac{l_\theta}{1-l_\theta},
\end{equation}
which indicates that $\Phi'(1)$ is real and $\Phi'(1)\in (-\infty, -1)\cup (0, +\infty)$.  

It can be verified that Eqs. (\ref{equ:azimuthal_slip_C_1}) has a special solution
\begin{equation}
\Phi=\dfrac{r^n}{\lambda},
\end{equation}
and its corresponding homogeneous differential equation is
\begin{equation}\label{equ:azimuthal_slip_C_1_homo}
r^2\Phi''+r\Phi'-(n^2+\lambda r^2)\Phi=0.
\end{equation}
From the theory of ordinary differential equation, this equation has two linearly independent solutions in $(0, 1]$. One of the two solutions can be represented as a generalized power series 
\begin{equation}\label{equ:series_solution_of_phi1}
\Phi_1=\sum_{m=0}^{\infty}B_mr^{m+\rho} \hspace{2mm}(B_0\neq 0),
\end{equation}
in which it can be obtained that $\rho=n$, $B_{2k+1}=0$ and $B_{2k}=\left(\dfrac{\lambda}{4}\right)^k\dfrac{B_0}{k!(n+k)!}$ using the standard undetermined coefficient method. Denoting $a_n=B_0$, Eqs. (\ref{equ:series_solution_of_phi1}) can be written as
\begin{equation}
\Phi_1=a_n r^n\sum_{k=0}^{\infty} \left(\dfrac{\lambda}{4}\right)^k\dfrac{r^{2k}}{k!(n+k)!}.
\end{equation}  
The other solution of Eqs. (\ref{equ:azimuthal_slip_C_1_homo}) has the following form
\begin{equation}
\Phi_2=\Phi_1\int_1^r\dfrac{1}{\Phi_1^2(s)}\exp{\left(-\int_1^s\dfrac{1}{t}\text{d}t\right)}\text{d}s=\Phi_1\int_1^r\dfrac{1}{s\Phi_1^2(s)}\text{d}s.
\end{equation}
However, by L'Hopital rule,
\begin{equation}
\lim_{r\to 0}\Phi_2(r)=\lim_{r\to 0}\dfrac{\int_1^r\dfrac{1}{s\Phi_1^2(s)}\text{d}s}{\dfrac{1}{\Phi_1(r)}}=\lim_{r\to 0}\dfrac{\dfrac{1}{r\Phi_1^2(r)}}{\dfrac{\Phi_1'(r)}{\Phi_1^2(r)}}=\infty,
\end{equation}
which is unphysical, and therefore $\Phi_2$ should not appear in the general solution of Eqs. (\ref{equ:azimuthal_slip_C_1}), i.e. the general solution of Eqs. (\ref{equ:azimuthal_slip_C_1}) can be solved as
\begin{equation}
\Phi=\dfrac{1}{\lambda}r^n+a_n r^n\sum_{k=0}^{\infty}\left(\dfrac{\lambda}{4}\right)^k\dfrac{r^{2k}}{k!(n+k)!}.
\end{equation}
For simplicity, denoting $\mu=\dfrac{\lambda}{4}$ and using the boundary condition $\Phi(1)=0$, one can solve for $a_n$ as
\begin{equation}
a_n=-\dfrac{1}{4\mu}\left(\sum_{k=0}^{\infty}\dfrac{\mu^k}{k!(n+k)!}\right)^{-1},
\end{equation}
consequently,
\begin{align}
\Phi'(1)=&\dfrac{n}{4\mu}+a_n\sum_{k=0}^{\infty}\dfrac{\mu^k(n+2k)}{k!(n+k)!} \notag \\
=&-\dfrac{1}{4\mu}\left(\sum_{k=0}^{\infty}\dfrac{\mu^k}{k!(n+k)!}\right)^{-1}\sum_{k=0}^{\infty}\dfrac{2k\mu^k}{k!(n+k)!},
\end{align}
i.e. $\mu$ satisfies
\begin{equation}\label{equ:phi1_mu}
\sum_{k=1}^{\infty}\dfrac{k\mu^{k-1}}{k!(n+k)!}+2\Phi'(1)\sum_{k=0}^{\infty}\dfrac{\mu^k}{k!(n+k)!}=0.
\end{equation}
In the following, we prove that $\mu$ has to be real and $\mu<0$ given Eqs. (\ref{equ:phi1_mu}). For simplicity, let $s=\Phi'(1)$ 
and define $f(z)$ as
\begin{equation}\label{def:fz}
f(z)=\sum_{k=0}^{\infty}\dfrac{z^k}{k!(n+k)!},
\end{equation}
where $z$ is complex. Then, Eqs. (\ref{equ:phi1_mu}) states that $\mu$ is a root of the equation $f'(z)+2sf(z)=0$. Note that
\begin{equation}
\left(z^{n+1}f'(z)\right)'=\left(\sum_{k=1}^{\infty}\dfrac{kz^{n+k}}{k!(n+k)!}\right)'=z^n\sum_{k=1}^{\infty}\dfrac{z^{n-k}}{(k-1)!(n+k-1)!}=z^n\sum_{k=0}^{\infty}\dfrac{z^k}{k!(n+k)!}=z^nf(z).
\end{equation}
Then, defining $f_\mu(z)=f(\mu z)$, it can be verified that
\begin{equation}\label{equ:fmu}
(z^{n+1}f_\mu'(z))'=\mu z^nf_\mu(z),
\end{equation}
in which the prime denotes the derivative with respect to $z$. Further, note that $\bar \mu$ is also a root of Eqs. (\ref{equ:phi1_mu}), because the coefficients are all real. That is to say,
\begin{equation}\label{equ:fmu_bar}
(z^{n+1}f_{\bar\mu}'(z))'=\bar \mu z^nf_{\bar\mu}(z),
\end{equation}
where $f_{\bar\mu}=f(\bar\mu z)$. 
Then, the difference between Eqs (\ref{equ:fmu}) multiplied by $f_{\bar\mu}(z)$ and Eqs. (\ref{equ:fmu_bar}) multiplied by $f_{\mu}(z)$, integrated along the real axis from 0 to 1 gives that
\begin{align}
& \int_0^1(z^{n+1}f_\mu'(z))'f_{\bar\mu}(z)\text{d}z-\int_0^1(z^{n+1}f_{\bar\mu}'(z))'f_\mu'(z)\text{d}z \notag \\
& = z^{n+1}f_\mu'(z)f_{\bar\mu}(z)\bigg|_0^1 -\int_0^1z^{n+1}f_\mu'(z)\text{d}f_{\bar\mu}(z)-z^{n+1}f_{\bar\mu}'(z)f_\mu(z)\bigg|_0^1+\int_0^1z^{n+1}f_{\bar\mu}'(z)\text{d}f_\mu(z) \notag \\
& = \int_0^1z^{n+1}|f_\mu'(z)|^2\text{d}z - \int_0^1z^{n+1}|f_\mu'(z)|^2\text{d}z + f_\mu'(1)f_{\bar\mu}(1) -f_\mu(1)f_{\bar\mu}'(1) \notag \\
& = 2s(\bar\mu-\mu)|f_\mu(1)|^2 \notag =(\mu-\bar\mu)\int_0^1z^n|f_\mu(z)|^2\text{d}z,
\end{align}
where the condition $f'(z)+2sf(z)=0$ is used to derive $f_\mu'(z)+2\mu sf_\mu(z)=0$ and $f_{\bar\mu}'(z)+2\bar\mu sf_{\bar\mu}(z)=0$. Then we have
\begin{equation}\label{equ:mu_minus_mubar}
(\mu-\bar\mu)\left(\int_0^1z^n|f_\mu(z)|^2\text{d}z+2s|f_\mu(1)|^2\right)=0.
\end{equation}

Similarly, the sum of Eqs (\ref{equ:fmu}) multiplied by $f_{\bar\mu}(z)$ and Eqs. (\ref{equ:fmu_bar}) multiplied by $f_{\mu}(z)$, integrated along the real axis from 0 to 1 gives that
\begin{align}\label{equ:mu_plus_mubar}
(\mu+\bar\mu)\left(\int_0^1z^n|f_\mu(z)|^2\text{d}z+2s|f_\mu(1)|^2\right)=-2\int_0^1z^{n+1}|f_\mu'(z)|^2\text{d}z,
\end{align}
which indicates $\int_0^1z^n|f_\mu(z)|^2\text{d}z+2s|f_\mu(1)|^2\neq 0$ because the right hand side is non-zero. Consequently, Eqs. (\ref{equ:mu_minus_mubar}) indicates that $\mu-\bar\mu=0$, i.e. $\mu$ is real.

Subsequently, we can deduce that $\mu<0$ if $s\in(0, \infty)$ because the term in the parentheses and the integral on the right hand side  of Eqs. (\ref{equ:mu_plus_mubar}) are all positive. In case of $s\in(-\infty, -1)$, if $\mu$ were positive, one would obtain
\begin{equation}
-2s\sum_{k=0}^{\infty}\dfrac{\mu^k}{k!(n+k)!}=\sum_{k=1}^{\infty}\dfrac{k\mu^{k-1}}{k!(n+k)!}=\sum_{k=0}^{\infty}\dfrac{\mu^k}{k!(n+k+1)!}\leqslant \sum_{k=0}^{\infty}\dfrac{\mu^k}{k!(n+k)!},
\end{equation}
and consequently $-2s\leqslant 1$, which would conflict with $s\in(-\infty, -1)$. Therefore, $\mu<0$. Finally, we obtain that $\mu<0$ for \BS{$s\in (-\infty, -1)\cup (0, +\infty)$}, i.e. for any value of $l_\theta\neq 1$. Since we have shown before that $\lambda<0$ for $l_\theta=1$ and for $l_\theta=0$, now we reach the conclusion that $\lambda$ is real and $\lambda<0$ for any $l_\theta\in [0, +\infty)$, regardless of $l_x$, i.e. the flow is rigorously linearly stable to perturbations with $\alpha=0$ with or without velocity slip.

\subsection{Analytical solution of the eigenvalue and eigenvector for $\alpha=0$ modes}
\label{sec:eigenvalue_eigenvector_alpha0}
We consider the general case with both streamwise and azimuthal slip. 
For $\Phi\not\equiv 0$, if $C\neq 0$, we obtain from Eqs. (\ref{equ:phi1_mu}) that $\mu=\dfrac{\lambda}{4}$ satisfies
\begin{equation}
(1-l_\theta)\sum_{k=1}^{\infty}\dfrac{k\mu^{k-1}}{k!(n+k)!}+2l_\theta\sum_{k=0}^{\infty}\dfrac{\mu^k}{k!(n+k)!}=0,
\end{equation}
where Eqs. (\ref{equ:Phi_prime_at_wall}) is used. The Bessel functions of integer order $n$ and $n+1$ read
\begin{equation}\label{Bessel_n}
J_n(z)=\sum_{k=0}^{\infty}\dfrac{(-1)^k}{k!(n+k)!}\left(\dfrac{z}{2}\right)^{2k+n}=\left(\dfrac{z}{2}\right)^n\sum_{k=0}^{\infty}\dfrac{1}{k!(n+k)!}\left(-\dfrac{z^2}{4}\right)^k,
\end{equation}
\begin{equation}\label{Bessel_nplus1}
J_{n+1}(z)=\sum_{k=0}^{\infty}\dfrac{(-1)^k}{k!(n+1+k)!}\left(\dfrac{z}{2}\right)^{2k+n+1}=\left(\dfrac{z}{2}\right)^{n+1}\sum_{k=0}^{\infty}\dfrac{1}{k!(n+k+1)!}\left(-\dfrac{z^2}{4}\right)^k.
\end{equation}
Denoting $\mu=-\dfrac{\eta^2}{4}$, i.e. the eigenvalue $\lambda=4\mu=-\eta^2$, it can be observed that $\eta$ is a root of the equation
\begin{equation}\label{equ:eta_being_root}
(1-l_\theta)J_{n+1}(z)+l_\theta z J_n(z)=0.
\end{equation}

Next, we show that $C\neq 0$ if $l_\theta\neq 1$. Assuming $C=0$ and $l_\theta\neq 1$, $\Phi'(1)+\Phi''(1)=0$ and the boundary condition $\Phi'(1)+l_\theta\Phi''(1)=0$ would give $\Phi'(1)=\Phi''(1)=0$. Recall that the solution to the homogeneous equation Eqs. (\ref{equ:azimuthal_slip_C_1_homo}) is 
\begin{equation}
\Phi_1=a_n r^n\sum_{k=0}^\infty \left(\dfrac{\lambda}{4}\right)^k\dfrac{r^{2k}}{k!(n+k)!},
\end{equation}
where $a_n$ is a constant.
Using the notation of (\ref{def:fz}), $f(\mu)=\sum_{k=0}^\infty\mu^k\dfrac{1}{k!(n+k)!}=0$ results from $\Phi_1(1)=0$ (note that $\Phi=\Phi_1$ if $C=0$), which would indicate that the corresponding $\eta$ satisfies $J_n(z)=0$. 
Further, $\Phi'(1)=0$ gives
\begin{equation}
\sum_{k=0}^\infty\mu^k\dfrac{n+2k}{k!(n+k)!}=0.
\end{equation}
In combination with $f(\mu)=0$, we would obtain that
\begin{equation}
\sum_{k=1}^\infty\mu^k\dfrac{1}{(k-1)!(n+k)!}=\sum_{k=0}^\infty\mu^k\dfrac{1}{k!(n+k+1)!}=0,
\end{equation}
which means that $\eta$ would also be a zero of $J_{n+1}(z)$, i.e. $\eta$ would be a zero of both $J_n(z)$ and $J_{n+1}(z)$. This would conflict with the fact that there exists no comment zero of $J_n(z)$ and $J_{n+1}(z)$. Therefore, $C\neq 0$ and $\eta$ is a root of Eqs. (\ref{equ:eta_being_root}) if $l_\theta\neq 1$.

We have proved before that $C=0$ if $l_\theta=1$, which gives $\Phi=\Phi_1$. Consequently, $\Phi(1)=\Phi_1(1)=0$ gives $f(\mu)=0$, which means $\eta$ is a root of $J_n(z)=0$, i.e. $\eta$ is also a root of Eqs. (\ref{equ:eta_being_root}) (note that the first term disappears if $l_\theta=1$). Therefore, $\eta$ is a root of Eqs. (\ref{equ:eta_being_root}) for any $l_\theta\geqslant 0$.

For the case of $\Phi\equiv 0$, Eqs. (\ref{equ:Phi_always_0_2}) and the corresponding boundary condition $\Omega(1)+l_x \Omega'(1)=0$ (see Eqs. (\ref{equ:BC2_2}) and (\ref{equ:BCx_alpha0})) imply that the eigenvector $\Omega$ has the same form as the solution $\Phi_1$, and we can deduce that $\lambda=-\gamma^2$, in which $\gamma$ is a root of
\begin{equation}\label{equ:gamma_being_root}
(1+nl_x)J_n(z)-l_x z J_{n+1}(z)=0.
\end{equation} 
For an eigenvalue $\lambda=-\eta^2$, the corresponding eigenvector can be solved as
\begin{equation}\label{eigenvector_phi_phi_ne0}
\Phi=J_n(\eta r)-J_n(\eta)r^n,
\end{equation}
\begin{equation}\label{eigenvector_omega_phi_ne0}
\Omega=b_nJ_n(\eta r)+\frac{2i}{(1+4l_x)n^2}\left(\frac{r}{2\eta}J_{n+1}(\eta r)-\frac{J_n(\eta)}{\eta^2}r^n\right),
\end{equation}
where $b_n$ is a constant and should be determined by the boundary condition Eqs. (\ref{equ:BCx_alpha0}). For an eigenvalue $\lambda=-\gamma^2$, the eigenvector can be solved as
\begin{equation}\label{eigenvector_phi0}
\Phi\equiv 0, \hspace{2mm} \Omega=J_n(\gamma r).
\end{equation}

\begin{figure}
\centering
\includegraphics[width=0.78\textwidth]{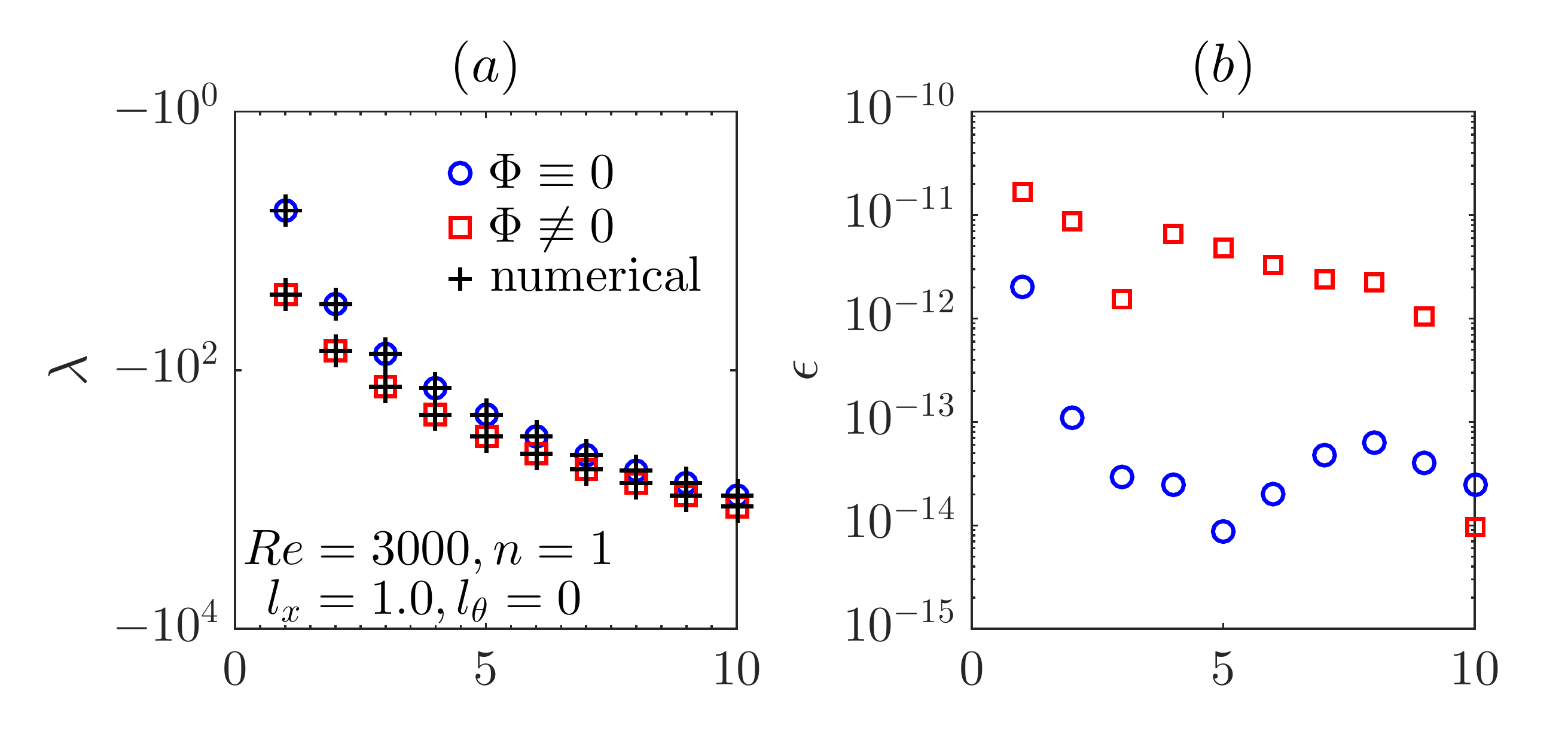}
\caption{\label{fig:eigenvalue_validation} Validation of the analytical eigenvalues against the numerical calculation for the case of $Re=3000$, $n=1$, $l_x=1.0$ and $l_\theta=0$. In (a), the first 20 eigenvalues are shown as squares and circles, and the numerical results are shown as crosses. The circles are the first ten eigenvalues (in descending order) for the cases with $\Phi\equiv 0$ and the squares are the first 10 eigenvalues (in descending order) for the $\Phi\not\equiv 0$ case. (b) The relative error $\epsilon$ between the analytical and numerical ones.}
\end{figure}

To sum up, there are always two groups of eigenvalues, corresponding to $\Phi\equiv 0$ (given by Eqs. \ref{equ:gamma_being_root}) and $\Phi\not\equiv 0$ (given by Eqs. \ref{equ:eta_being_root}), respectively. Particularly, for the no-slip case, Eqs. (\ref{equ:eta_being_root}) reduces to $J_{n+1}(z)=0$ and Eqs. (\ref{equ:gamma_being_root}) reduces to $J_n(z)=0$, and it is known that the zeros of $J_n(z)$ and $J_{n+1}(z)$ distribute alternately. Therefore, in the no-slip case, these two groups of eigenvalues distribute alternately either. For the streamwise slip case, Eqs. (\ref{equ:eta_being_root}) still reduces to $J_{n+1}(z)=0$, i.e. the $\Phi\not\equiv 0$ eigenvalues do not change with $l_x$, whereas the $\Phi\equiv 0$ eigenvalues will change with $l_x$. However, there cannot be common roots between $J_{n+1}(z)=0$ and Eqs. {\ref{equ:gamma_being_root}}, otherwise there would be common zeros between $J_n(z)$ and $J_{n+1}(z)$, which conflicts with the fact that there are none common zero between the two. Therefore, as $l_x$ changes, the two groups of eigenvalues distribute in the same alternating pattern as in the no-slip case and there is no over-taking between the two groups, see Figure \ref{fig:spectrum_streamwise_slip}(a) and Figure \ref{fig:eigenvalue_validation}(a). However, this behavior is not guaranteed in the azimuthal slip case as there can be common roots between Eqs. (\ref{equ:eta_being_root}) and (\ref{equ:gamma_being_root})) given $l_x=0$ and $l_\theta=1.0$, i.e. the roots of $J_n(z)=0$. Nonetheless, it should be noted that the common roots can only exist at $l_\theta=1.0$. This implies that, when eigenvalues change with $l_\theta$, an over-taking between the two groups may occur at precisely $l_\theta=1.0$.

\begin{figure}
\centering
\includegraphics[width=0.82\textwidth]{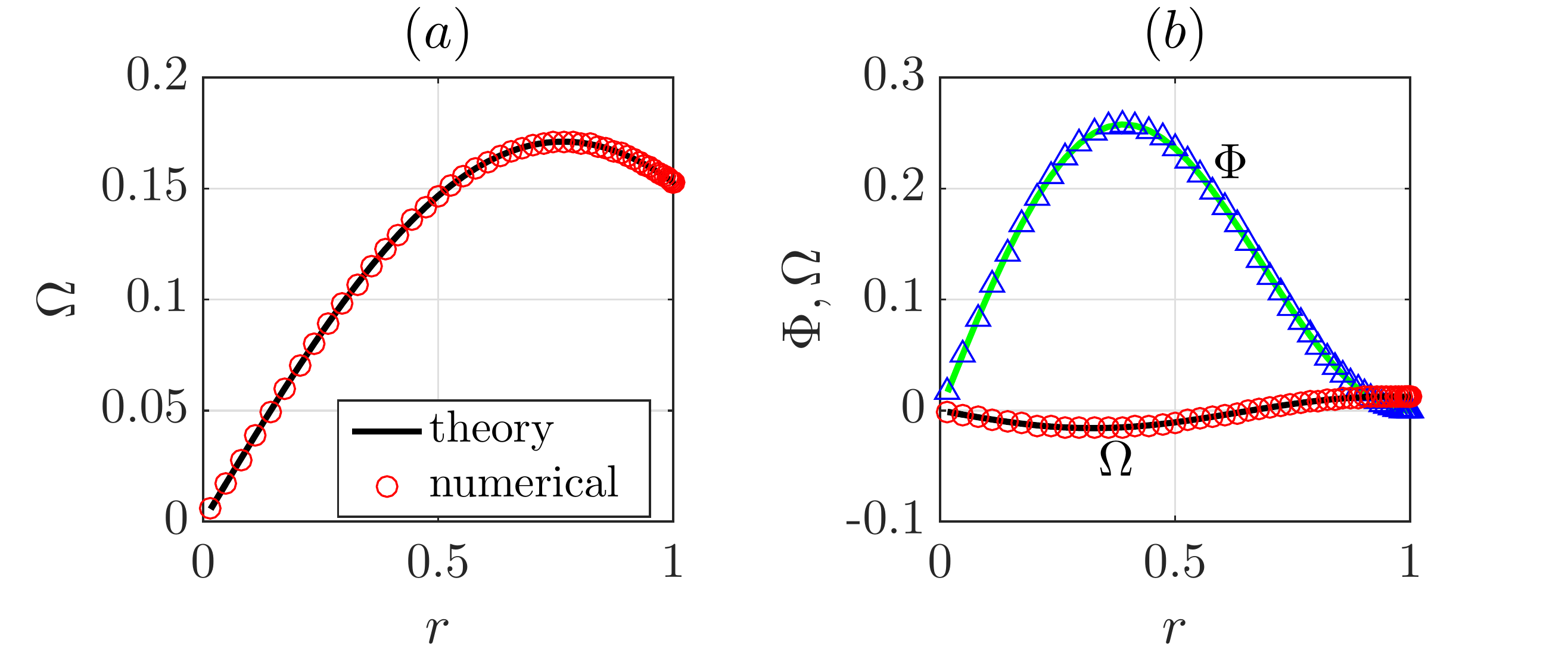}
\caption{\label{fig:eigenvectors_validation} Validation of the analytical eigenvectors against the numerical calculation for the case of $Re=3000$, $n=1$, $l_x=1.0$ and $l_\theta=0$. (a) The $\Omega$ component of the eigenvector associated with the leading eigenvalue (the leftmost blue circle in Figure \ref{fig:eigenvalue_validation}($a$)). The $\Phi$ component is zero and is not shown. (b) The $\Omega$ and $\Phi$ component of the eigenvector associated with the second largest eigenvalue (the leftmost red square in Figure \ref{fig:eigenvalue_validation}($a$)).}
\end{figure}

Figure \ref{fig:eigenvalue_validation} shows the comparison between our analytical solution  of the two groups of eigenvalues and numerical calculation for the streamwise slip case of $Re=3000$, $n=1$, $l_x=1.0$ and $l_\theta=0$. In panel (a), blue circles are analytical solutions of the first 10 largest eigenvalues given by Eqs. (\ref{equ:gamma_being_root}), i.e. the corresponding eigenvectors all have $\Phi\equiv 0$, and red squares represent the first 10 largest eigenvalues given by Eqs. (\ref{equ:eta_being_root}), i.e. the corresponding eigenvectors all have $\Phi\not\equiv 0$. Clearly, the leading eigenvalue is and will always be associated with $\Phi\equiv 0$ disturbances because no over-taking between the two groups of eigenvalues can occur as $l_x$ varies, as we concluded in Section \ref{sec:eigenvalue_eigenvector_alpha0}. These analytical solutions agree very well with the numerical calculations (the crosses) with relative errors of $\mathcal{O}(10^{-11})$ or lower, see panel (b).  
The eigenvector associated with the leading eigenvalue (the leftmost circle in Figure \ref{fig:eigenvalue_validation}(a)) is plotted in Figure \ref{fig:eigenvectors_validation}(a). The black line shows the analytical solution given by (\ref{eigenvector_phi0}) and the circles show the numerical calculation. The $\Phi$ part of the eigenvector is not shown because $\Phi\equiv 0$. Figure \ref{fig:eigenvectors_validation}(b) shows the eigenvector associated with the second largest eigenvalue (the leftmost red square in Figure \ref{fig:eigenvalue_validation}(a)), which has a non-zero $\Phi$ part. The figure shows that, for both $\Phi$ and $\Omega$, our analytical solutions (lines) agree very well with the numerical calculations (symbols). This comparison validates our theory about the eigenvalue and eigenvector.

\begin{figure}
\centering
\includegraphics[width=0.85\textwidth]{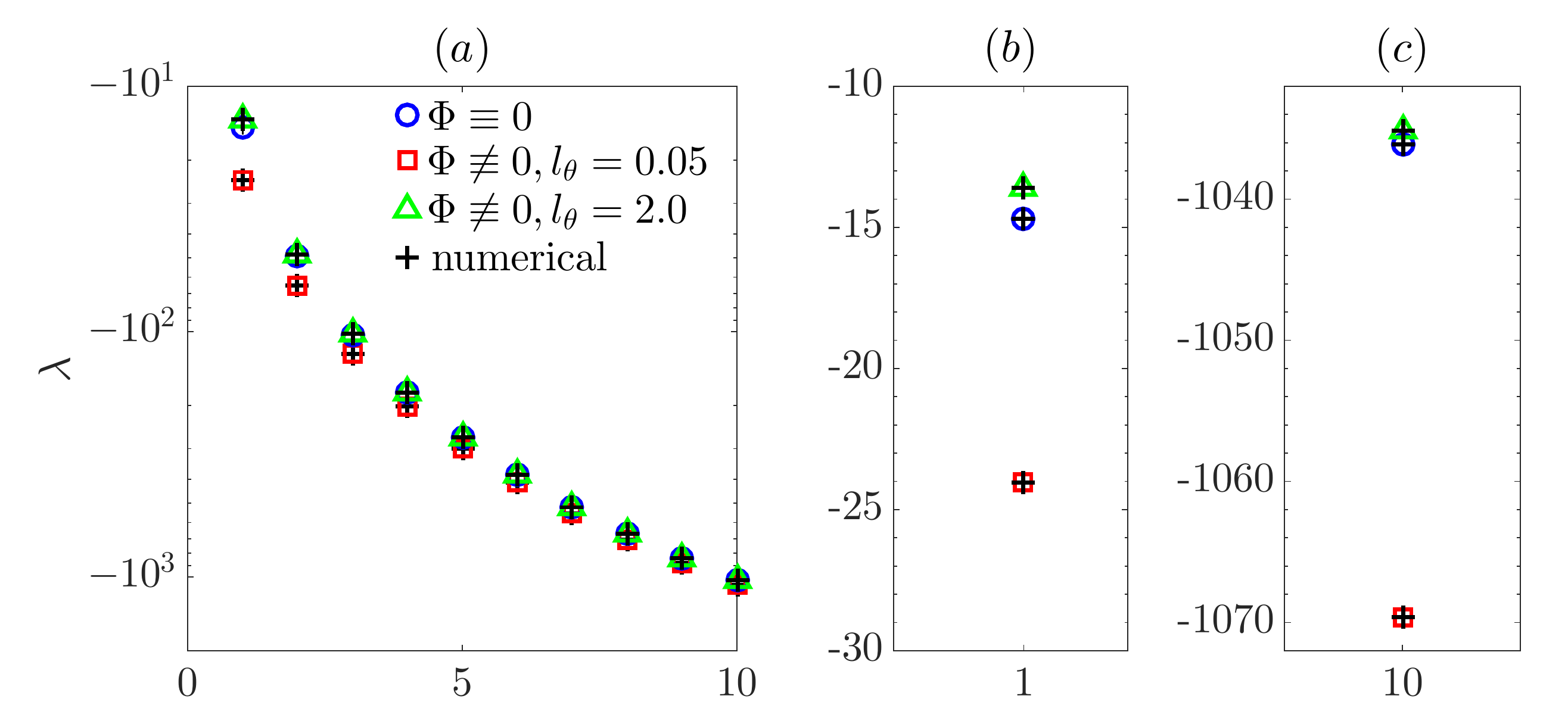}
\caption{\label{fig:eigenvalue_validation_azimuthal_slip} Eigenvalues for $Re=3000$, $n=1$ and $l_x=0$ with $l_\theta=0.05$ and 2.0. In (a), analytical eigenvalues are shown as circles ($\Phi\equiv 0$, for which eigenvalues are independent of $l_\theta$), squares ($\Phi\not\equiv 0$ and $l_\theta=0.05$) and triangles ($\Phi\not\equiv 0$ and $l_\theta=2.0$), and the numerical calculations are shown as crosses. Panel (b) and (c) show the zoom-in of the two ends of the spectrum shown in (a).}
\end{figure}

The two groups of eigenvalues of the azimuthal slip cases of $l_\theta=0.05$ and 2.0 for $Re=3000$, $n=1$ and $l_x=0$ are also shown in Figure \ref{fig:eigenvalue_validation_azimuthal_slip}. Again, perfect agreement between the analytical and numerical ones is observed. We can see that, for $l_\theta=0.05$, the $\Phi\not\equiv 0$ group is entirely below the $\Phi\equiv 0$ group, which is independent of $l_\theta$, whereas is entirely above the $\Phi\equiv 0$ group for $l_\theta=2.0$, indicating that an over-taking indeed occurs between the two groups as $l_\theta$ increases. Therefore, for $l_\theta<1.0$, the leading eigenvalue is associated with $\Phi\equiv 0$ disturbances and does not change with $l_\theta$ (see also Figure \ref{fig:spectrum_spanwise_slip}(a)), whereas it is associated with $\Phi\not\equiv 0$ disturbances and increases with $l_\theta$ for $l_\theta>1.0$.

\subsection{The dependence of the leading eigenvalue on slip length for $\alpha=0$ modes}
\label{sec:dependence_on_slip_length_alpha0}

Denoting $F(z,l_\theta)=(1-l_\theta)J_{n+1}(z)+l_\theta zJ_n(z)$, it can be obtained that, as $z\to 0$,
\begin{equation}\label{equ:F(z,ltehta)_small_z}
J_n(z)\sim\dfrac{z^n}{2^n n!},\hspace{2mm} F(z,l_\theta)\sim\left(\dfrac{1-l_\theta}{2^{n+1}(n+1)!}+\dfrac{l_\theta}{2^n n!}\right)z^{n+1}.
\end{equation}
It can be seen that $\dfrac{1-l_\theta}{2^{n+1}(n+1)!}+\dfrac{l_\theta}{2^n n!}>0$ for $l_\theta\geqslant 0$, therefore, $F(z,l_\theta)$ is positive for sufficiently small $z$. Let $z_1$ be the minimum root of $F(z,l_{\theta 1})=0$ and $z_2$ be the minimum root of $F(z,l_{\theta 2})=0$. If $l_{\theta 1}<l_{\theta 2}$, it can be derived that
\begin{align}\label{equ:Fz1theta2_being_negative}
F(z_1,l_{\theta 2})=&(1-l_{\theta 2})J_{n+1}(z_1)+l_{\theta 2} z_1J_n(z_1) \notag \\
=& (1-l_{\theta 2})J_{n+1}(z_1)-\dfrac{1-l_{\theta 1}}{l_{\theta 1}}l_{\theta 2}J_{n+1}(z_1) \notag \\
=& \dfrac{l_{\theta 1}-l_{\theta 2}}{l_{\theta 1}}J_{n+1}(z_1)<0.
\end{align}
In (\ref{equ:Fz1theta2_being_negative}), $J_{n+1}(z)>0$ follows from that, at the minimum positive zero of $J_{n+1}(z)$, denoted as $z_0$, we have $F(z_0,l_{\theta 1})<0$ because $J_n(z_0)<0$. We showed before that $F(z,l_{\theta 1})>0$ at sufficiently small $z$, therefore, the minimum positive zero of $F(z,l_{\theta 1})$, $z_1$, should be smaller than $z_0$ given that $F(z,l_{\theta 1})$ is continuous with respect to $z$, i.e. $z_1<z_0$, and therefore $J_{n+1}(z_1)>0$. Consequently, given $F(z_1,l_{\theta 2})<0$, there must be a zero in $(0, z_1)$, i.e. $z_2<z_1$ because the function $F(z,l_{\theta 2})$ is continuous with respect to $z$. This states that, for the case of $\Phi\not\equiv 0$, the maximum eigenvalue $\lambda$, denoted as $\lambda_1$ in the following, increases as $l_\theta$ increases and is independent of $l_x$. Similarly, one can deduce that the minimum root of Eqs. (\ref{equ:gamma_being_root}) decreases as $l_x$ increases, consequently, the maximum eigenvalue for the $\Phi\equiv 0$ case, denoted as $\lambda_2$, increases as $l_x$ increases and is independent of $l_\theta$.
\begin{figure}
\centering
\includegraphics[width=0.85\textwidth]{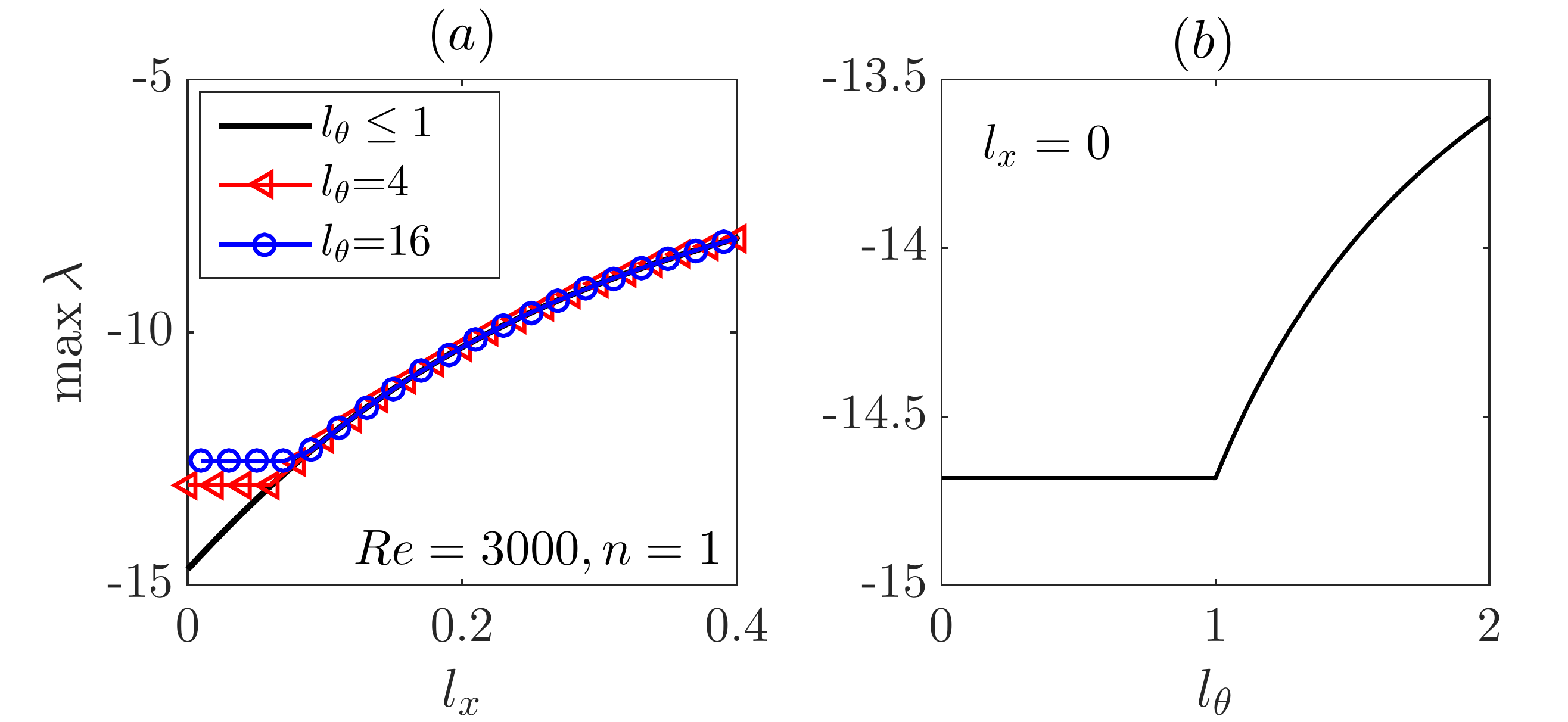}
\caption{\label{fig:eigval_dependence_on_slip_length} The dependence of $\max{\lambda}$ on slip length for $Re=3000$ and $n=1$. (a) Both streamwise and azimuthal slip are present. The black line shows the dependence on $l_x$ given $l_\theta \leqslant 1$. Symbol lines show two cases for $l_\theta > 1$. (b) The dependence on $l_\theta$ in case of $l_x=0$.}
\end{figure}
For the special case of $l_x=0$, Eqs. (\ref{equ:gamma_being_root}) becomes $J_n(z)=0$ and for the case of $l_\theta=1$, Eqs. (\ref{equ:eta_being_root}) turns into $zJ_n(z)=0$. 
Clearly, these two cases share the non-zero roots, i.e. $\lambda_1=\lambda_2$. Therefore, the minimum root of Eqs. (\ref{equ:gamma_being_root}) is always greater than that of Eqs. (\ref{equ:eta_being_root}), i.e. $\lambda_1>\lambda_2$, when $l_\theta<1$. 
This explains why, for a given $l_\theta\leqslant 1$, $\max{\lambda}$ increases monotonically as $l_x$ increases from zero, whereas for a given $l_\theta>1$, $\max{\lambda}$ first stays constant and only starts to increase until $l_x$ is increased above a threshold, see Figure \ref{fig:eigval_dependence_on_slip_length}(a). If only azimuthal slip is present, i.e. $l_x=0$, $\max{\lambda}$ firstly stays constant and only starts to increase precisely at $l_\theta=1$, see Figure \ref{fig:eigval_dependence_on_slip_length}(b). The data shown in the inset of Figure \ref{fig:spectrum_spanwise_slip}(a) also support this conclusion, see that $\max{\lambda}$ for $l_\theta=0.005$, 0.05 and 0.5 are identical. It can also be inferred that, given a fixed $l_x>0$ and that $\lambda_2$ increases with $l_x$, $\max{\lambda}$ can only start to increase as $l_\theta$ increases at some $l_\theta>1$.  

In summary, the maximum eigenvalue of $\alpha=0$ modes is an increasing function of $l_\theta$ or $l_x$ (may not be strictly increasing, depending on the slip length setting, as Figure \ref{fig:eigval_dependence_on_slip_length} shows) and is independent of the Reynolds number, which is obvious as $Re$ does not appear in Eqs. (\ref{equ:eta_being_root}) and (\ref{equ:gamma_being_root}). Nevertheless, the eigenvalues remain negative.

\section{Non-modal stability}
It has been known that in many shear flows (e.g., pipe, channel and plane-Couette flows), small disturbances can be transiently amplified due to the non-normality of the linearized equations, despite their asymptotic linear stability \citep{Schmid1994,Meseguer2003,Schmid2007}. This transient amplification is believed to play an important role in the subcritical transition in shear flows. Here,
we also investigated the effects of the anisotropic slip on the non-modal stability of the flow. The same method for calculating the transient growth described by \citet{Schmid1994} is adopted. The transient growth at time $t$ for a mode $(\alpha, n)$ is defined as
\begin{equation}\label{equ:TG_definition}
G(t;\alpha,n)=\underset{{E(0)\neq 0}}{\text{max}}\frac{E(t)}{E(0)},
\end{equation}
where $E(t)=\int_V\boldsymbol u(t)^2 dV$ is the kinetic energy of the perturbation $\boldsymbol u$ integrated in the whole flow domain at time $t$. For linearly stable flow, $G$ will reach the maximum, $G_{\text max}$, at certain time and monotonically decay at larger times. For linearly unstable flow, $G$ can be either non-monotonic or monotonic at early stages, depending on the competition between the modal and non-modal growth, and will be dominated by the exponential growth of the most unstable disturbance at large times.

\begin{figure}
\centering
\includegraphics[width=0.9\textwidth]{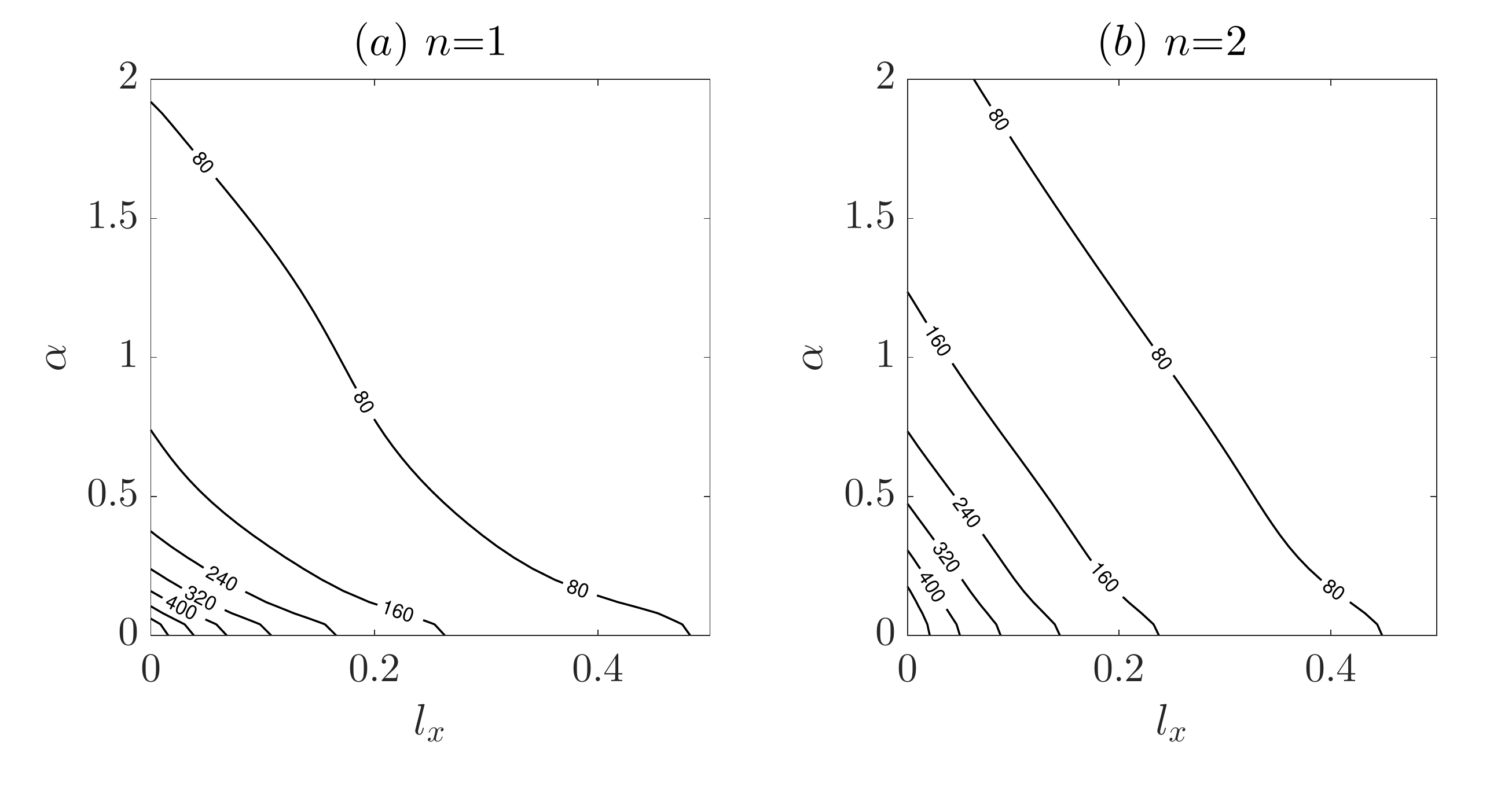}
\caption{\label{fig:nonmodal_streamwise_slip} The maximum transient growth, $G_{\text max}$, at $Re=3000$ plotted in the $l_x$-$\alpha$ plane for $n=1$ (a) and $n=2$ (b). Azimuthal slip length $l_\theta$=0. The contour level step is 80 in both panels.}
\end{figure}

\begin{figure}
\centering
\includegraphics[width=0.9\textwidth]{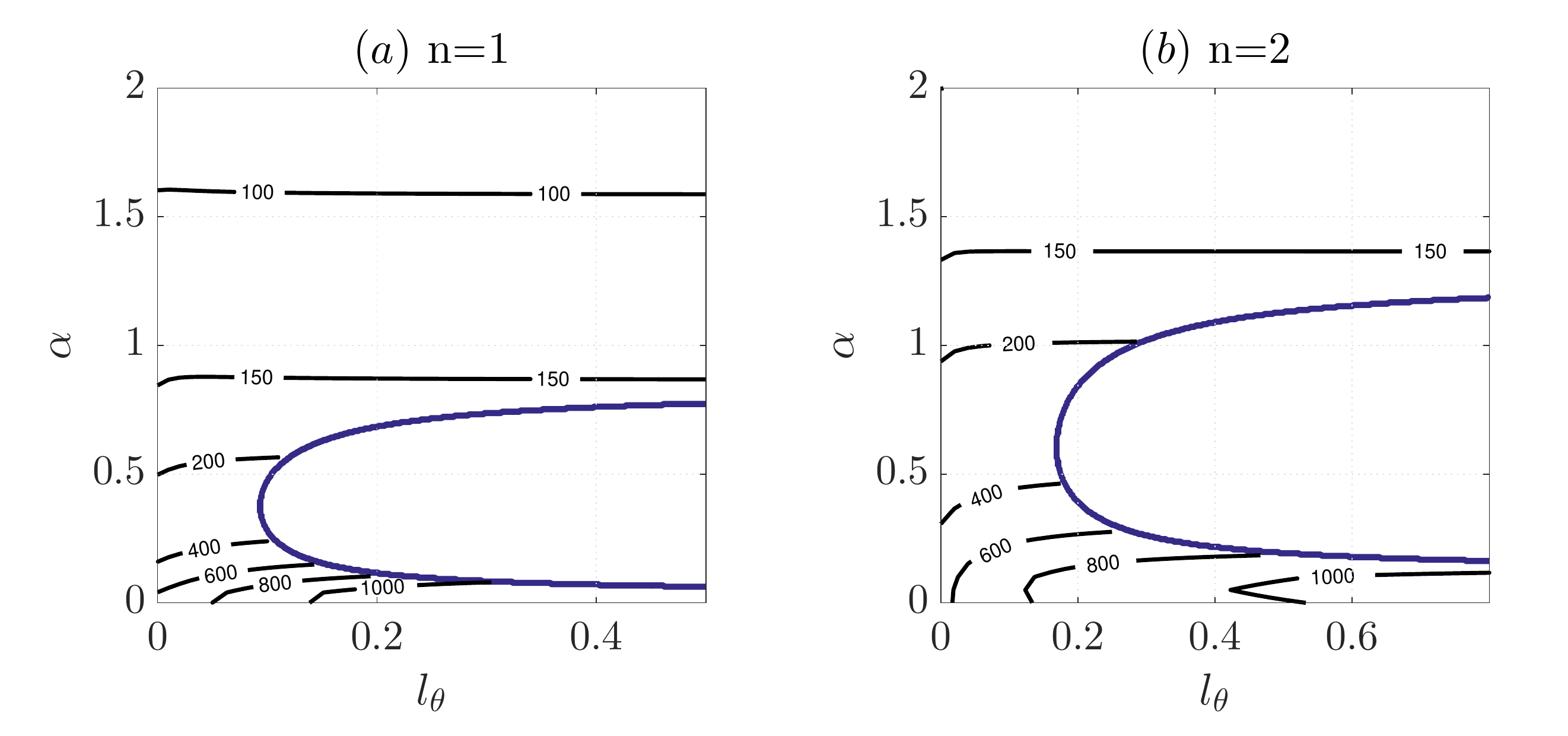}
\caption{\label{fig:nonmodal_spanwise_slip} The maximum transient growth, $G_{\text max}$, at $Re=3000$ plotted in the $l_\theta$-$\alpha$ plane for $n=1$ (a) and $n=2$ (b). Streamwise slip length $l_x$=0. The bold lines enclose the linearly unstable regions.}
\end{figure}

Figure \ref{fig:nonmodal_streamwise_slip} shows $G_{\text max}$ at $Re=3000$ in the $l_x$-$\alpha$ plane ($l_x=0$) for $n=1$ and $n=2$ (low azimuthal wavenumbers are generally most amplified by non-normality). From the contour lines we can see that steamwise slip reduces $G_{\text max}$ and the decrease is monotonic as $l_x$ increases. Intuitively, streamwise slip reduces the background shear such that the lift-up mechanism \citep{Brandt2014} is subdued. Therefore, the transient growth should be reduced as our results show. The most amplified mode is still the $(\alpha,n)=(0,1)$ mode (streamwise rolls) as in the no-slip case \citep{Schmid1994,Meseguer2003}.

Figure \ref{fig:nonmodal_spanwise_slip} shows $G_{\text max}$ for the azimuthal slip case at $Re=3000$ in the $l_\theta$-$\alpha$ plane. Azimuthal wavenumbers $n=1$ and $n=2$ are considered. From the orientation of the contour lines we can see that azimuthal slip increases $G_{\text max}$. Presumably, azimuthal slip can enhance streamwise vortices because it reduces wall friction and allows finite azimuthal velocity at the wall, and therefore, the lift-up mechanism can be enhanced exhibiting increased transient growth as our results show. For $n=1$, in the $l_\theta$ range investigated, the most amplified mode is still the $(\alpha,n)=(0,1)$ mode as in the no-slip case. However, for $n=2$, as $l_\theta$ increases, the most amplified mode is no longer the streamwise independent one but one with a small finite streamwise wavenumber (long wavelength), see panel (b). For example, the most amplified mode is approximately $\alpha=0.05$ for $l_\theta$ above about 0.1. Similar behavior has also been reported for channel flow \citep{Chai2019}. The two bold lines in the figure enclose the linearly unstable regions in which $G_{\text max}$ is theoretically infinite, therefore, the linearly unstable region is left blank. Unlike the streamwise slip case where the slip length significantly reduces the transient growth throughout the $\alpha$ and $l_x$ ranges investigated (see Figure \ref{fig:nonmodal_streamwise_slip}), azimuthal slip only significantly affects the transient growth for small $\alpha$ and nearly does not affect that of larger $\alpha$, see the nearly horizontal contour lines for relatively large $\alpha$. Besides, even for small $\alpha$, $G_{\text max}$ quickly saturates as $l_\theta$ increases. To sum up, azimuthal slip only affects the transient growth of the modes with small streamwise wavenumbers and the effect saturates as the slip length increases. 

\begin{figure}
\centering
\includegraphics[width=0.99\textwidth]{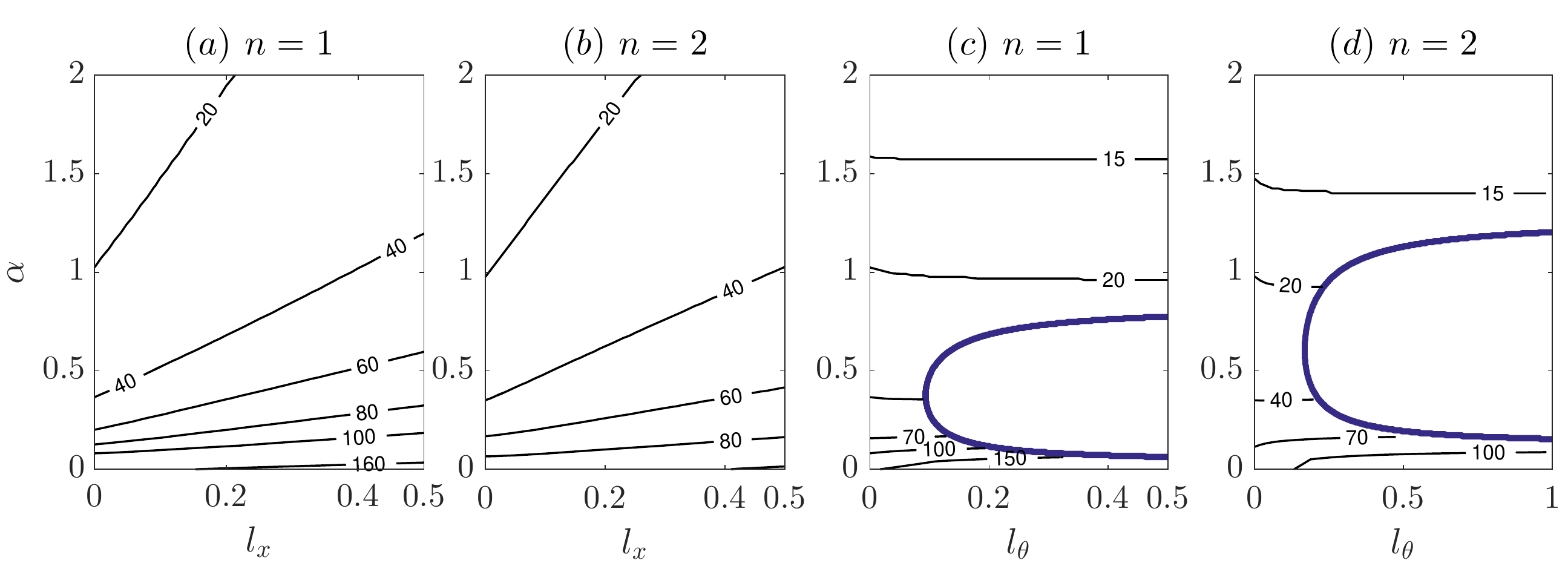}
\caption{\label{fig:nonmodal_tmax} The time instant when $G_{\text max}$ is reached, $t_{\text max}$, in the $l_x$-$\alpha$ plane for $n=1$ (a) and $n=2$ (b) and in the $l_\theta$-$\alpha$ plane for $n=1$ (c) and $n=2$ (d).}
\end{figure}

Figure \ref{fig:nonmodal_tmax} shows the time instant when $G_{\text max}$ is reached, $t_{\text max}$, for the cases shown in Figure \ref{fig:nonmodal_streamwise_slip} and \ref{fig:nonmodal_spanwise_slip}. In the streamwise slip case (panel (a,b)), for both $n=1$ and 2, the slip increases $t_{\text max}$ and the effect is more significant for larger $\alpha$. In the azimuthal slip case (panel (c,d)), for small $\alpha$ ($\lesssim 0.2$), $t_{\text max}$ is also slightly increased by the slip, whereas for larger $\alpha$, $t_{\text max}$ is slightly decreased by the slip, in contrast to the streamwise slip case. Overall, for the modes with small $\alpha$, i.e. most amplified modes due to non-normality, both streamwise and azimuthal slip only mildly increase $t_{\text max}$.

\begin{figure}
\centering
\includegraphics[width=0.9\textwidth]{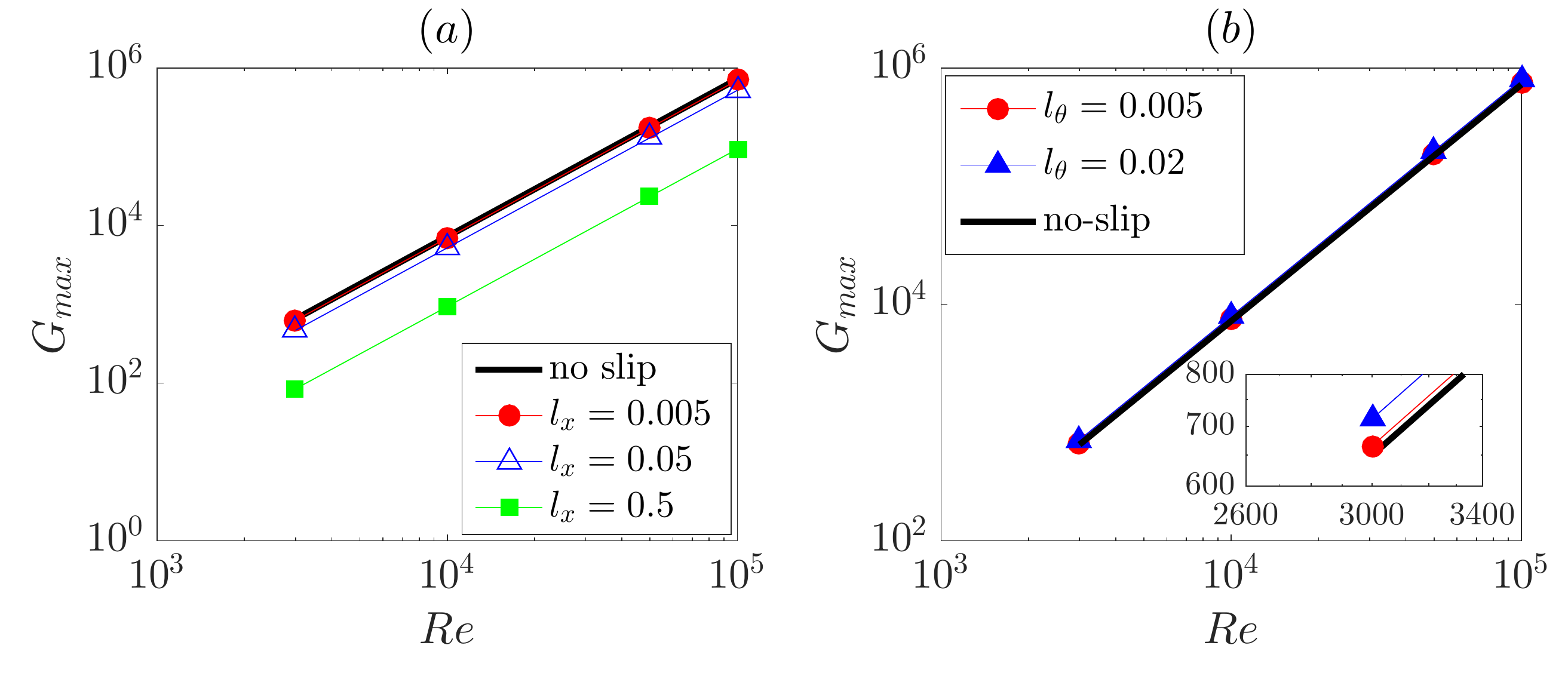}
\caption{\label{fig:nonmodal_Re_scaling} The scaling of global $G_{\text max}$ with $Re$. (a) Streamwise slip $l_x=0.005$, 0.05 and 0.5. (b) Azimuthal slip $l_\theta=0.005$ and 0.02. In both panels, the bold black line shows the scaling for the no-slip case. The inset in (b) is a zoom-in around $Re=3000$.}
\end{figure}

In the no-slip case, the global $G_{\text max}$ (maximized over $\alpha$ and $n$) is known to scale as $Re^2$ when $Re$ is large \citep{Meseguer2003}. In the presence of streamwise and azimuthal slip, the scaling is also investigated and shown in Figure \ref{fig:nonmodal_Re_scaling}. In the streamwise slip case (panel (a)), the lines for $l_x=0.005$, 0.05 and 0.5 appear to be parallel to the no-slip case with a downward vertical shift, suggesting a $Re^2$-scaling for any streamwise slip length. Similarly, for azimuthal slip, the scaling also seem to be $Re^2$. For this case, we only investigated $l_\theta=0.005$ and 0.02 in the linearly stable regime because larger $l_\theta$ will cause linear instability.

\section{Conclusions and discussions}
It has been well established that the classic pipe flow is (asymptotically) linearly stable. In this paper, we studied the effect of velocity slip on the linear stability of pipe flow. 
Our results show that the leading eigenvalue increases with streamwise slip length $l_x$ but remains negative, i.e. streamwise slip renders the flow less stable but does not cause linear instability, similar to the effect of isotropic slip length on the flow \citep{Prusa2009}. Interestingly, our results suggest that the leading eigenvalue is independent of $Re$, or equivalently, the slowest decay rate of disturbances scales as $Re^{-1}$ (note that time is scaled by $Re^{-1}$ in our formulation). It should be pointed out that this scaling holds 
at sufficiently high Reynolds numbers ($\gtrsim 10^4$). For relatively low Reynolds numbers (100 and 1000 in our study), there is a very slight deviation from the scaling for $l_x\lesssim 0.1$ and the deviation is substantial at larger $l_x$.
The $Re^{-1}$-scaling of the decay rate is the same as what was observed for the mode $(\alpha,n)=(0,1)$ of the classic pipe flow by \citet{Meseguer2003}. Besides, our results show that the streamwise wavenumber at which the eigenvalue maximizes is not $\alpha=0$ but also scales as $Re^{-1}$. However, if $l_x$ is very large ($\gtrsim 1.0$, and note that in applications the slip length is generally much smaller), the eigenvalue maximizes at $\alpha=0$ at relatively low Reynolds numbers (100 and 1000 in our study) and this scaling also only holds at high Reynolds numbers ($\gtrsim 10^4$).

This destabilizing effect appears to be opposite to the stabilizing effect of streamwise slip reported for channel flow (the stabilizing effect is mainly observed for 2-D perturbations) \citep{Lauga2005,Min2005,Chai2019}. Here, we only provide a possible explanation for the least stable/most unstable perturbation (referred to as the leading perturbation for simplicity) of the two flows. We speculate that the different flow structures of the leading perturbations of the two flows are responsible. For pipe flow, we proved that the $u_r$ component of the leading perturbation for $\alpha=0$ modes vanishes. The globally leading perturbation has very small $\alpha$ ($\sim \frac{1}{Re}$), which indicates that $u_r$ should be nearly vanishing. Therefore, the production rate of kinetic energy ($-\int_Vu_ru_x\frac{\text{d}U_x}{\text{d}r}dV$) should be also very small and the decay rate of disturbances should be dominated by the dissipation rate. Intuitively, velocity slip reduces the dissipation rate due to the reduced wall friction, therefore, the decay rate of the least stable perturbation decreases, i.e. the flow appears to be destabilized. In contrast, for channel flow, the leading perturbations are 2-D Tollmien-Schlichting waves for small slip length, which have a substantial wall-normal velocity component (comparable to the streamwise component).  
Therefore, the kinetic energy production is significant and even dominant in the variation of the kinetic energy. Streamwise slip reduces the base shear ($\frac{\text{d}U_x}{\text{d}r}$) and therefore subdues the production. If the reduction in the production rate outweighs the decrease in the energy dissipation rate due to the reduced wall friction, the flow will be stabilized. This is probably why stabilizing effect of the streamwise slip on channel flow was observed.

On the contrary, azimuthal slip, given sufficiently large slip length, causes linear instability, similar to the finding of \citet{Chai2019} for channel flow. Our results show that azimuthal slip destabilizes helical waves with wavelengths considerably larger than the pipe diameter, whereas it does not affect the stability of waves with much shorter wavelengths and in the long wavelength limit, i.e. $\alpha\to 0$. The critical Reynolds number decreases sharply as $l_\theta$ increases and gradually levels off at around a few hundred as $l_\theta\gtrsim 0.3$ and at approximately 260 as $l_\theta\to \infty$. Similar destabilizing effect was reported for channel flow \citep{Chai2019}. Azimuthal slip serves as an example for a perturbation to the linear operator associated with the linearized Navier-Stokes equations with no-slip boundary condition that destabilizes the originally stable system.

Regarding the stability of the classic pipe flow to streamwise independent perturbations, using an energy analysis, \citet{Joseph1971} concluded the absolute and global stability of the flow, i.e. the flow is asymptotically (as $t\to \infty$) stable to such perturbations with arbitrary amplitude. Here, for the linear case and from a mathematical point of view, we rigorously proved that the eigenvalues of streamwise independent modes ($\alpha=0$) are real and negative, for arbitrary slip length and arbitrary Reynolds number. Besides, the eigenvalue of the $\alpha=0$ modes is proved to be strictly independent of Reynolds number in our formulation, in agreement with the numerical calculation by \citet{Meseguer2003}. We derived analytical solutions to the eigenvalue and eigenvector for $\alpha=0$ modes and verified our theory by numerical calculations.
We also proved that, the eigenvalues of $\alpha=0$ modes consist of two groups: One group is associated with disturbances with $\Phi\equiv 0$, i.e. $u_r\equiv 0$, and the other is associated with disturbances with $\Phi\not\equiv 0$, i.e. $u_r\not\equiv 0$ (see Fig \ref{fig:eigenvalue_validation} and Figure \ref{fig:eigenvalue_validation_azimuthal_slip}). The two groups distribute alternately. For the streamwise slip case, the latter group stays constant while the former group changes with $l_x$. It is the other way round for the azimuthal slip case. Interestingly, for the streamwise slip case, the leading eigenvalue belongs to the $\Phi\equiv 0$ group and does not switch group as $l_x$ changes, whereas for the azimuthal slip case, it switches from the $\Phi\equiv 0$ group to the $\Phi\not\equiv 0$ group as $l_\theta$ crosses 1.0 from below (see Figure \ref{fig:eigenvalue_validation_azimuthal_slip}). When both $l_x$ and $l_\theta$ are non-zero, $l_x$ dominates the leading eigenvalue if $l_\theta<1$. If $l_\theta>1$, the leading eigenvalue first stays constant and can only start to increase at a threshold as $l_x$ increases (see Figure \ref{fig:eigval_dependence_on_slip_length}). Such analytical solutions might inspire asymptotic analysis in the limit of small streamwise wavenumber. 

Non-modal analysis shows that streamwise slip greatly reduces the transient growth, whereas azimuthal slip significantly increases the transient growth for disturbances with very small steamwise wavenumbers but nearly does not affect that for disturbances with larger streamwise wavenumbers, aside from the linear instability caused by the slip. Both streamwise slip and azimuthal slip give the $Re^2$-scaling of the maximum transient growth, the same as in the no-slip case \citep{Schmid1994,Meseguer2003}. Similar effects were observed for channel flow \citep{Chai2019}.

Linear instability caused by anisotropic slip at low Reynolds numbers is of interest for small flow systems, such as microfluidics, in which the Reynolds number is usually low but the non-dimensional slip length can be significantly large using advanced surface texturing techniques. The instability can be exploited to enhance mixing or heat transfer in applications involving small flow systems. Larger non-modal growth caused by azimuthal slip can potentially cause earlier subcritical transition to turbulence. Besides, introducing modal instability into originally sub-critical flows may also help to better understand the transition mechanism in such flows.

\section{Acknowledgements}
The authors acknowledge financial support from the National Natural Science Foundation of China under grant number 91852105 and 91752113 and from Tianjin University under grant number 2018XRX-0027. The comments from the reviewers are also highly appreciated.
\section{Declaration of interests}

The authors report no conflict of interest.

\appendix
\section{Numerics}
\label{sec:numerics}
Firstly, we briefly explain the implementation of the boundary condition (\ref{equ:BC2_1}) and (\ref{equ:BC2_2}) in the eigenvalue problem for the linear system (\ref{equ:NS2}).

The eigenvalue equation reads
\begin{equation}
\label{equ:eigen_equation_2}
-L^{-1}M{\boldsymbol q}=\lambda {\boldsymbol q},
\end{equation}
where $q$ is the unknown vector composed of $\hat\Phi$ and $\hat\Omega$, see (\ref{def:q}). Boundary conditions (\ref{equ:BC2_1}) and (\ref{equ:BC2_2}) couple $\hat\Phi$ and $\hat\Omega$, unlike in the no-slip case. In our Fourier-Fourier-Chebyshev collocation discretization, $\boldsymbol q$ is a $2N\times 1$ vector and the operator $-L^{-1}M$ is discretized as a $2N\times 2N$ matrix, where $N$ is the number of collocation grid point on the radius. Adopting the Chebyshev differentiation matrix of \citet{Trefethen2000} which is of spectral accuracy, the matrix is dense and the radial differentiation of $\boldsymbol q$ at a single grid point is calculated using the value of $\boldsymbol q$ at all collocation points on the radius. Given that $\hat\Phi$ is known at $r=1$, i.e. $\hat\Phi=0$ at $r=1$, the size of the system can be reduced by one. The boundary conditions (\ref{equ:BC2_1}) and (\ref{equ:BC2_2}) give two linear algebraic equations about $\hat\Phi$ and $\hat\Omega$ at the collocation points, from which we can further eliminate two unknowns. By doing so, the system size is reduced by 3, i.e. to $(2N-3)\times (2N-3)$, from which the eigenvalue problem can be solved with the boundary conditions being taken accounted for.

Secondly, we show the convergence test of our numerical calculation. We consider the case of $Re=3000$, $\alpha=0.5$ and $n=1$, as presented in Figure \ref{fig:spectrum_streamwise_slip}(b) and Figure \ref{fig:spectrum_spanwise_slip}(b). We change the number of Chebyshev points and check the convergence of the mean-mode, wall-mode and center-mode separately. Table \ref{tab:streamwise_slip}, \ref{tab:spanwise_slip} and \ref{tab:ltheta_infinity} shows the resolutions and the eigenvalue of an arbitrarily selected mean mode and the rightmost eigenvalues corresponding to the wall- and center-mode. It can be seen that grid numbers $N=32$, 64 and 128 give very close values of the eigenvalues, which differ only after about 7 digits after the decimal point (the relative difference is $\mathcal{O}(10^{-11})$), for all the slip length of 0.005, 0.05, 0.5 and the case of $l_\theta=\infty$. The convergence test show that, for the calculation of the rightmost eigenvalues and for calculating the mean mode at $Re=3000$, 32 points are sufficient. Solely for calculating the rightmost eigenvalue, we used 32 points for $Re=100$, 1000 and 10000, and 64 points for $Re=10^5$ and $Re=10^6$. We checked the convergence by doubling the number of grid points and found these numbers sufficient. It should be noted that, although 32 points are sufficient for calculating both of the rightmost eigenvalue and the spectrum at $Re=3000$, more grid points may be needed for accurately calculating the spectrum than for calculating the rightmost eigenvalue at higher Reynolds numbers \citep{Trefethen2000,Meseguer2003}.


\begin{table}
\centering
$Re=3000$, $\alpha=0.5$, $n=1$, $l_x=0.005$ and $l_\theta=0$
\begin{tabular}{ccc}
\hline
$N$	& mean-mode	& wall-mode  \\
32		& -622.180970763082-1006.803438258427$i$	& 
-106.901701568167-600.000309196652$i$   \\
64		& -622.180970617437-1006.803438460110$i$	& -106.901701568129-600.000309195248$i$   \\
128		& -622.180970549502-1006.803438403115$i$	& -106.901701574121-600.000309193566$i$  \\
\hline
$N$ & center-mode & \\
32  &  -87.144443682906-1299.140360156589$i$ & \\
64  &  -87.144443682943-1299.140360156482$i$ & \\
128 &  -87.144443681739-1299.140360155874$i$ & \\
\hline
\end{tabular}
$Re=3000$, $\alpha=0.5$, $n=1$, $l_x=0.05$ and $l_\theta=0$
\begin{tabular}{ccc}
\hline
$N$	& mean-mode	& wall-mode  \\
32		& -530.922788817816-953.518793751342$i$	& 
-68.522846170133-646.557599422173$i$   \\
64		& -530.922788815748-953.518793737632$i$	& 
-68.522846166147-646.557599425783$i$   \\
128		& -530.922788808102-953.518793754673$i$	& 
-68.522846166147-646.557599425783i  \\
\hline
$N$ & center-mode & \\
32  &  -80.309678415026-1203.366029301252$i$ & \\
64  &  -80.309678415070-1203.366029301237$i$ & \\
128 &  -80.309678413737-1203.366029300673$i$ & \\
\hline
\end{tabular}

$Re=3000$, $\alpha=0.5$, $n=1$, $l_x=0.5$ and $l_\theta=0$
\begin{tabular}{ccc}
\hline
$N$	& mean-mode	& wall-mode  \\
32		& -582.2041335894776-833.8681859837275$i$	& 
-32.3669400581618-546.6565442837639$i$   \\
64		& -582.2041335893674-833.8681859850334$i$	& 
-32.3669400570418-546.6565442835596$i$   \\
128		& -582.2041336044291-833.8681859835294$i$	& 
-32.3669400643051-546.6565442852506$i$  \\
\hline
$N$ & center-mode & \\
32  &  -36.8382964038017-757.7599724649968$i$ & \\
64  &  -36.8382964035955-757.7599724652722$i$ & \\
128 &  -36.8382963977050-757.7599724664044$i$ & \\
\hline
\end{tabular}

\caption{\label{tab:streamwise_slip}The convergence of the eigenvalue corresponding to the mean mode (arbitrarily selected) and the rightmost wall-mode and center-mode as the radial grid number $N$. The streamwise slip cases of $l_x=0.005$, 0.05 and 0.5 for $Re=3000$, $\alpha=0.5$, $n=1$ and $l_\theta=0$ are listed.}
\end{table}

\begin{table}
\centering
$Re=3000$, $\alpha=0.5$, $n=1$, $l_\theta=0.005$ and $l_x=0$
\begin{tabular}{ccc}
\hline
$N$	& mean-mode	& wall-mode  \\
32		& -878.143428857252-1000.225260064803$i$	& 
-147.212095816072-565.187041064729$i$  \\
64		& -878.143429638119-1000.225258436485$i$	& 
-147.212095816866-565.187041057138$i$   \\
128		& -878.143429646590-1000.225258451813$i$	& 
-147.212095817673-565.187041057973$i$  \\
\hline
$N$ & center-mode & \\
32  &  -88.016026669448-1311.995868654545$i$ & \\
64  &  -88.016026669520-1311.995868654469$i$ & \\
128 &  -88.016026668333-1311.995868653794$i$ & \\
\hline
\end{tabular}
$Re=3000$, $\alpha=0.5$, $n=1$, $l_\theta=0.05$ and $l_x=0$
\begin{tabular}{ccc}
\hline
$N$	& mean-mode	& wall-mode  \\
32		& -818.018362966476-992.813175584590$i$	& 
-30.701258513947-434.347551571943$i$   \\
64		& -818.018361688704-992.813172416195$i$	& 
-30.701258511491-434.347551570607$i$   \\
128		& -818.018361708603-992.813172372158$i$	& 
-30.701258513706-434.347551572422$i$  \\
\hline
$N$ & center-mode & \\
32  &  -88.015855528115-1312.001178576785$i$ & \\
64  &  -88.015855528186-1312.001178576701$i$ & \\
128 &  -88.015855526922-1312.001178575825$i$ & \\
\hline
\end{tabular}

$Re=3000$, $\alpha=0.5$, $n=1$, $l_\theta=0.5$ and $l_x=0$
\begin{tabular}{ccc}
\hline
$N$	& mean-mode	& wall-mode  \\
32		& -794.161410360041-1006.311883474201$i$	& 
33.866513836957-391.619593748286$i$   \\
64		& -794.161408826090-1006.311880389063$i$	& 
33.866513839707-391.619593747032$i$   \\
128		& -794.161408715927-1006.311880293383$i$	& 
33.866513840046-391.619593749138$i$  \\
\hline
$N$ & center-mode & \\
32  &  -88.013747237500-1312.003625793072$i$ & \\
64  &  -88.013747237549-1312.003625792987$i$ & \\
128 &  -88.013747236338-1312.003625792171$i$ & \\
\hline
\end{tabular}

\caption{\label{tab:spanwise_slip}The convergence of the eigenvalue corresponding to the mean mode (arbitrarily selected) and the rightmost wall-mode and center-mode as the radial grid number $N$. The azimuthal slip cases of $l_\theta=0.005$, 0.05 and 0.5 for $Re=3000$, $\alpha=0.5$, $n=1$ and $l_x=0$ are listed.}
\end{table}

\begin{table}
\centering
$Re=3000$, $\alpha=0.5$, $n=1$, $l_\theta=\infty$ and $l_x=0$
\begin{tabular}{ccc}
\hline
$N$	& mean-mode	& wall-mode  \\
32		& -637.771735030362-1019.911762553870$i$	& 
45.558397965700-383.081240915272$i$  \\
64		& -637.771735090255-1019.911762536828$i$	& 
45.558397968779-383.081240913500$i$   \\
128		& -637.771735157437-1019.911762419706$i$	& 
45.558397965363-383.081240919212$i$  \\
\hline
$N$ & center-mode & \\
32  &  -88.013312066798-1312.003902287461$i$ & \\
64  &  -88.013312066887-1312.003902287400$i$ & \\
128 &  -88.013312039270-1312.003902226469$i$ & \\
\hline
\end{tabular}

\caption{\label{tab:ltheta_infinity}The convergence of the eigenvalue corresponding to the mean mode (arbitrarily selected) and the rightmost wall-mode and center-mode as the radial grid number $N$. The azimuthal slip case of $l_\theta=\infty$ for $Re=3000$, $\alpha=0.5$, $n=1$ and $l_x=0$ is listed.}
\end{table}

\section{The dependence of the leading eigenvalue on the azimuthal wavenumber $n$ for $\alpha=0$}
\label{sec:dependence_on_n_alpha0}

Our numerical calculations in Section \ref{sec:streamwise_slip} and \ref{sec:azimuthal_slip} showed that $n=1$ modes are the least stable/most unstable modes. Here we show the dependence of $\max{\lambda}$ of $\alpha=0$ modes on the azimuthal wavenumber $n$ and prove that $n=1$ is indeed the least stable azimuthal mode. For this purpose, we only need to prove that the minimum non-zero roots of Eqs. (\ref{equ:eta_being_root}) and (\ref{equ:gamma_being_root}) all increase with $n$.

Note that the root of Eqs. (\ref{equ:eta_being_root}) is independent of $l_x$. Because the zeros of $J_n(z)$ and $J_{n+1}(z)$ distribute alternately, it can be easily seen that, if $l_\theta\leqslant 1$, the minimum positive root of Eqs. (\ref{equ:eta_being_root}) is located between the minimum positive zeros of $J_n(z)$ and $J_{n+1}(z)$. Therefore, the minimum positive root of Eqs. (\ref{equ:eta_being_root}) increases with $n$ because the positive zeros of $J_n(z)$ and $J_{n+1}(z)$ all increase with $n$, i.e., $\lambda_1$ decreases as $n$ increases if $l_\theta\leqslant 1$.

If $l_\theta>1$, we need to prove that the minimum positive root of 
\begin{equation}\label{equ:eta_being_root_2}
g_n(z)=(1-l_\theta)J_n(z)+l_\theta zJ_{n-1}(z)=0, \hspace{2mm} n\geqslant 2
\end{equation}
is smaller than that of $g_{n+1}(z)=0$, i.e. Eqs. (\ref{equ:eta_being_root}). 
We already showed in Eqs. (\ref{equ:F(z,ltehta)_small_z}) that $F(z, l_\theta)>0$ at sufficiently small $z$, i.e. $g_n(z)>0$ for sufficiently small $z$. Denoting the minimum positive root of Eqs. (\ref{equ:eta_being_root}) as $z_0$, we only need to show that $g_n(z_0)<0$. Using the property of Bessel function of
\begin{equation}\label{equ:property_bessel_function}
J_{n+1}(z)+J_{n-1}(z)=\frac{2n}{z}J_n(z)
\end{equation}
and Eqs. (\ref{equ:eta_being_root}), $g_n(z)$ can be rewritten as
\begin{equation}\label{equ:g(z0)_Jn(z0)}
g_n(z_0)=(1-l_\theta)J_n(z_0)+l_\theta z_0\left(\frac{2n}{z_0}-\frac{l_\theta}{l_\theta-1}z_0\right)J_n(z_0).
\end{equation}
It is easily seen that $J_n(z_0)>0$, therefore, we need to show that 
\begin{equation}
(1-l_\theta)+l_\theta z_0\left(\frac{2n}{z_0}-\frac{l_\theta}{l_\theta-1}z_0\right)<0,
\end{equation} 
or equivalently, 
\begin{equation}\label{equ:z0square}
z_0^2>(1-l_\theta^{-1})(2n-1+l_\theta^{-1}),
\end{equation} 
in order to prove that $g_n(z_0)<0$, given $l_\theta>1$. Noticing that $(1-l_\theta^{-1})(2n-1+l_\theta^{-1})<2n-1$ if $l_\theta>1$, we can prove $g_n(z_0)<0$ if we can show that $z_0^2>2n-1$.

In Section \ref{sec:dependence_on_slip_length_alpha0} we proved that the minimum positive root of Eqs. (\ref{equ:eta_being_root}) decreases as $l_\theta$ increases, therefore, $z_0$ is minimized at $l_\theta=+\infty$. To prove that $z_0^2>2n-1$ for any $l_\theta>1$, we only need to show that $z_0^2>2n-1$ holds for $l_\theta=+\infty$, with which Eqs. (\ref{equ:eta_being_root}) reduces to $J_{n+1}(z)-zl_\theta J_n(z)=0$. That $F(z,l_\theta)>0$ at sufficiently small $z$ indicates that $J_{n+1}(z)-zJ_n(z)<0$ at sufficiently small $z$ given $l_\theta=+\infty$. Therefore, $J_{n+1}(z_0)-z_0 J_n(z_0)=0$ requires that $J_{n+1}(z)-zJ_n(z)<0$, i.e. $J_{n+1}(z)<zJ_n(z)$ for any $z<z_0$. In fact, we can show that $J_{n+1}(z)<zJ_n(z)$ if $0<z^2<2n-1$ such that $z_0$ has to satisfy $z_0^2>2n-1$. Using the series form of Bessel function, $J_{n+1}(z)<zJ_n(z)$ means
\begin{equation}\label{equ:J(n+1)_zJn_ltheta_infty}
\sum_{k=0}^{+\infty}\frac{(-1)^2}{k!}\frac{1}{(n+k+1)!}\left(\frac{z}{2}\right)^{2k+n+1}<z\sum_{k=0}^{+\infty}\frac{(-1)^2}{k!}\frac{1}{(n+k)!}\left(\frac{z}{2}\right)^{2k+n}.
\end{equation}
Because of the absolute convergence of the two infinite series in Eqs. (\ref{equ:J(n+1)_zJn_ltheta_infty}), we only need to show that for any positive even number $k$, 
\begin{align}\label{equ:J(n+1)_zJn_ltheta_infty2}
\frac{1}{k!}\frac{1}{(n+k+1)!}\left(\frac{z}{2}\right)^{2k+n+1} &-\frac{1}{(k+1)!}\frac{1}{(n+k+2)!}\left(\frac{z}{2}\right)^{2k+2+n+1}< \notag \\
& \frac{z}{k!}\frac{1}{(n+k)!}\left(\frac{z}{2}\right)^{2k+n}-\frac{z}{(k+1)!}\frac{1}{(n+k+1)!}\left(\frac{z}{2}\right)^{2k+2+n},
\end{align}
if $0<z^2<2n-1$. Rearranging Eqs. (\ref{equ:J(n+1)_zJn_ltheta_infty2}), we have
\BS{
\begin{equation}\label{equ:J(n+1)_zJn_ltheta_infty3}
z^2<4(k+1)(n+k+2),
\end{equation}
}which obviously holds because $z^2<2n-1<4n+4k+8\leq4(k+1)(n+k+2)$. To sum up, we have proved that $J_{n+1}(z)<zJ_n(z)$ if $z^2<2n-1$, therefore, $z_0$, which satisfies $J_{n+1}(z_0)-z_0 J_n(z_0)=0$, must satisfy $z_0^2>2n-1$. Consequently, Eqs. (\ref{equ:z0square}) and $g_n(z_0)<0$ hold, and thusly the minimum positive root of Eqs. (\ref{equ:eta_being_root_2}) is smaller than $z_0$, which indicates that the minimum positive root of Eqs. (\ref{equ:eta_being_root}) increases with $n$, i.e. $\lambda_1$ decreases with $n$.

Next, we prove that the minimum positive root of Eqs. (\ref{equ:gamma_being_root}) also increases with $n$. The equation
\begin{equation}\label{equ:hn}
h_n(z)=(1+nl_x)z^nJ_n(z)-l_xz^{n+1}J_{n+1}(z)=0
\end{equation}
share the non-zero roots with Eqs. (\ref{equ:gamma_being_root}), therefore, we only need to prove the same statement for Eqs. (\ref{equ:hn}). Denoting the minimum positive zero of $h_{n-1}(z)$ as $z_0$, we next show that $h_{n}(z)$ monotonically increases in $[0, z_0]$ such that there is no positive root of Eqs. (\ref{equ:hn}) in $[0, z_0]$, i.e. the minimum positive root of Eqs. (\ref{equ:hn}) increases with $n$. Using the property of Bessel function of $(z^{n+1}J_{n+1}(z))'=z^{n+1}J_n(z)$, where `$'$' denotes the derivative with respect to $z$, we take the derivative of $h_n(z)$ with respect to $z$ and obtain
\begin{equation}\label{equ:hn_derivative}
h_n'(z)=(1+nl_x)z^nJ_{n-1}(z)-l_xz^{n+1}J_{n}(z)=z(h_{n-1}(z)+l_xz^{n-1}J_{n-1}(z)).
\end{equation}
It is easy to see that $h_{n-1}(z)$ is positive at sufficiently small $z$ (the derivation is similar to that of $F(z, l_\theta)$ being positive at sufficiently small $z$, see Eqs. (\ref{equ:F(z,ltehta)_small_z})), consequently, $h_{n-1}(z)> 0$ in $(0, z_0)$. As $z_0$ is smaller than the minimum positive zero of $J_{n-1}(z)$, we have $J_{n-1}(z)> 0$ in $(0, z_0)$. Therefore, $h_n'(z)> 0$ in $(0, z_0)$, i.e. $h_n(z)$ monotonically increases and there is no positive root of Eqs. (\ref{equ:hn}) in $(0,z_0]$. In other words, the minimum positive root of $h_{n}(z)=0$ is always larger than that of $h_{n-1}(z)=0$, i.e. the minimum positive root of Eqs. (\ref{equ:gamma_being_root}) increases with $n$, and therefore $\lambda_2$ decreases with $n$. Now, we reach the conclusion that $\max{\lambda}=\max\{\lambda_1,\lambda_2\}$ decreases with $n$ for $\alpha=0$ modes because $\lambda_1$ and $\lambda_2$ both decrease with $n$.
 
%

\end{document}